\documentclass{aa}  

\usepackage{graphicx,nicefrac}
\usepackage{txfonts}
\usepackage{hyperref}
\usepackage{color}
\usepackage{pdflscape}
\usepackage{multirow}
\usepackage[capitalise]{cleveref}

\DeclareRobustCommand{\ion}[2]{\textup{#1\,\textsc{\lowercase{#2}}}}
\DeclareRobustCommand{\ion}[2]{\textup{#1\,\textsc{\lowercase{#2}}}}
\DeclareRobustCommand{\teff}{T_{\mathrm{eff}}}
\DeclareRobustCommand{\logg}{\log g}
\DeclareRobustCommand{\mh}{\mathrm{[M/H]}}
\DeclareRobustCommand{\feh}{\mathrm{[Fe/H]}}
\DeclareRobustCommand{\alph}{[\alpha\mathrm{/M}]}
\DeclareRobustCommand{\alphfe}{[\alpha\mathrm{/Fe}]}
\DeclareRobustCommand{\vmic}{\varv_\mathrm{mic}}

\DeclareRobustCommand{\kms}{$\mathrm{km s}^{-1}$}

\mathchardef\mhyphen="2D

\begin{document}

   \title{Gaia FGK benchmark stars: Spectral library and abundances of $\alpha$ and Fe-peak elements of the  third version
}
\titlerunning{Gaia FGK benchmark stars v3: Spectral library and abundances.}
    \author{
L. Casamiquela\inst{\ref{LIRA}} \and C. Soubiran\inst{\ref{LAB}} \and  P. Jofr\'e\inst{\ref{UDP}} \and S. Vitali\inst{\ref{UDP}} \and S. Blanco-Cuaresma\inst{\ref{ADS},\ref{UDS}} \and N. Lagarde\inst{\ref{LAB}} \and D. Slumstrup\inst{\ref{GTC},\ref{IAC}}\and U. Heiter\inst{\ref{UAO}} \and J. T. Palmerio\inst{\ref{CEA}} \and N. Brouillet\inst{\ref{LAB}} \and  S. Elgueta\inst{\ref{UDP}} \and A. Rojas--Arriagada\inst{\ref{USC},\ref{Millenium},\ref{CIRAS}} \and C. Aguilera--G\'omez\inst{\ref{PUC}} \and I. Hernández--Araya\inst{\ref{PUC}} \and O. L. Creevey\inst{\ref{OCA}} \and L. Balaguer--Nú\~nez\inst{\ref{IEEC},\ref{ICC},\ref{UB}} \and R. Carrera\inst{\ref{BOL}}
}

\institute{
LIRA, Observatoire de Paris, Université PSL, Sorbonne Université, Université Paris Cité, CY Cergy Paris Université, CNRS, 92190 Meudon, France\label{LIRA}
\email{laia.casamiquela@obspm.fr}
\and
Laboratoire d'Astrophysique de Bordeaux, Univ. Bordeaux, CNRS, B18N, all\'ee Geoffroy Saint-Hilaire, 33615 Pessac, France\label{LAB}
\and
Instituto de Estudios Astrof\'isicos, Facultad de Ingenier\'ia y Ciencias, Universidad Diego Portales, Av. Ej\'ercito Libertador 441, Santiago, Chile 
\label{UDP}
\and
Harvard-Smithsonian Center for Astrophysics, 60 Garden Street, Cambridge, MA 02138, USA\label{ADS}
\and
Faculty of Psychology, UniDistance Suisse, Brig, Switzerland\label{UDS}
\and
GRANTECAN, Cuesta de San José s/n, E-38712, Breña Baja, La Palma, ES \label{GTC}
\and
Instituto de Astrofísica de Canarias, E-38205 La Laguna, Tenerife, ES\label{IAC}
\and
Observational Astrophysics, Department of Physics and Astronomy, Uppsala University, Box 516, SE-751 20 Uppsala, Sweden\label{UAO}
\and
Université Paris-Saclay, Université Paris Cité, CEA, CNRS, AIM, 91191 Gif-sur-Yvette, France
\label{CEA}
\and
Departamento de F\'isica, Universidad de Santiago de Chile, Av. Victor Jara 3659, Santiago, Chile
\label{USC}
\and
Millennium Institute of Astrophysics, Avenida Vicuña Mackenna 4860, Macul, Santiago 82-0436, Chile
\label{Millenium}
\and
Center for Interdisciplinary Research in Astrophysics and Space Exploration (CIRAS), Universidad de Santiago de Chile, Santiago, Chile
\label{CIRAS}
\and
Instituto de Astrof\'isica, Pontificia Universidad Cat\'olica de Chile, Av. Vicu\~na Mackenna 4860, 782-0436 Macul, Santiago, Chile \label{PUC}
\and
Université Côte d'Azur, Observatoire de la Côte d'Azur, CNRS, Laboratoire Lagrange, Bd de l'Observatoire, CS 34229, 06304 Nice cedex 4, France
\label{OCA}
\and
Institut d'Estudis Espacials de Catalunya (IEEC), Esteve Terradas, 1, Edifici RDIT, Campus PMT-UPC, 08860 Castelldefels (Barcelona), Spain\label{IEEC}
\and
Institut de Ci\`encies del Cosmos (ICCUB), Universitat de Barcelona (UB), Mart\'i i Franqu\`es 1, E-08028 Barcelona, Spain\label{ICC}
\and
Departament de Física Qu\`antica i Astrof\'isica (FQA), Universitat de Barcelona (UB), Mart\'i i Franqu\`es 1, E-08028 Barcelona, Spain\label{UB}
\and
INAF-Osservatorio di Astrofisica e Scienza dello Spazio, via P. Gobetti 93/3, 40129, Bologna, Italy\label{BOL}
}
  
 \date{Received \today, accepted }

  \abstract
  {Accurate determination of the chemical abundances in stars plays a pivotal role in understanding stellar structure and evolution, nucleosynthesis, and the chemical enrichment history of the Milky Way. Benchmark stars with precise and accurate atmospheric parameters and abundances are indispensable for calibrating spectroscopic surveys and testing stellar atmosphere models.}
  {This study focuses on the compilation of high-quality spectra and the determination of chemical abundances of iron-peak and $\alpha$ elements for the third version of the Gaia FGK benchmark stars (GBSv3).}
  {We compiled spectra of the GBSv3 from public archives and complemented these with our observations. We used fundamental atmospheric parameters to perform a spectroscopic analysis using the public code iSpec and derived the chemical abundances.}
  {We compiled a homogeneous spectral library of high-resolution (42,000) and high signal-to-noise ($>100$) normalised spectra for 202 stars, including the 192 GBSv3, nine stars with indirect measurement of the angular diameter from previous Gaia FGK benchmark versions, and the Sun. Using four radiative transfer codes, we derived chemical abundances of 13 chemical species (Fe I, Fe II, Mg I, Si I, Ca I, Ti I, Ti II, Sc II, V I, Cr I, Mn I, Co I, Ni I). We performed an in-depth study of several sources of error.}
  {The GBSv3 contributes to the legacy samples of spectroscopic reference stars through improved statistics and homogeneity. This work offers the community a homogeneous spectral library and robust reference abundances for iron-peak and $\alpha$ elements, supported by an extensive analysis of the associated uncertainties.}


   \maketitle

\section{Introduction}
Over the past decade, significant observational efforts have been dedicated to uncovering the history of our Galaxy through the analysis of large samples of stars with abundances obtained from spectroscopic surveys.
A lesson learned from the first generation of high-resolution spectroscopic surveys of the Milky Way is that the abundance scales can differ significantly depending on the analysis methods \citep{Jofre+2017,Jonsson+2018,Worley+2020,Hegedus+2023}, implying that different scientific conclusions may be reached depending on which spectroscopic dataset is used.
It is then of paramount importance to have comprehensive validation samples that can put the different survey data onto the same scale, thus ensuring the data from all surveys are scientifically consistent \citep{Jofre+2019}. 
An effort has already been made to stitch together surveys via data-driven approaches \citep{Ho+2017,Wheeler+2020,Nandakumar2022,Thomas+2024}, such as The Cannon \citep{Ness+2015}, among others.
But large benchmark samples are still needed to ensure a robust homogenisation using such tools.
At the same time, these calibration samples are useful for a variety of purposes, including as science verification of new instruments \citep{Adibekyan+2020,Strassmeier+2018}, for the testing of systematic uncertainties in analysis techniques \citep{BlancoCuaresma2019}, for calibration of line lists \citep{Heiter+2021}, and as a means of pushing the limits of our understanding of stellar atmospheres, structure, and evolution \citep{Amarsi+2022,Creevey+2024}.

The Gaia FGK benchmark stars (GBS) were assembled as a key reference set for this purpose, and they provide robust, homogeneous determinations of stellar parameters \citep{Heiter+2015}; chemical compositions \citep{Jofre+2014, Jofre+2015, Hawkins+2016}; and high-quality spectral libraries \citep{BlancoCuaresma+2014}.
The GBS were selected to span F, G, and K (and M) spectral types across various luminosities and metallicities with sufficient observational data (particularly interferometry) to allow for determination of their effective temperature and surface gravity independently of spectroscopy, achieving a precision of 1–2\%.
The fact of having fundamental atmospheric parameters for these stars imposes strict constraints on spectroscopic analyses, enabling the determination of accurate reference values for chemical abundances.

The first GBS sample consisted of 30 stars \citep{Heiter+2015,Jofre+2014}, and the second version \citep{Jofre+2018} revised these stars and included metal-poor candidates analysed by \citet{Hawkins+2016} to construct a second set of 35 stars.
Recently, a third version of the GBS sample (GBSv3) has been presented by \citet{Soubiran+2024}, providing fundamental atmospheric parameters for a set of 192 stars.
Compared to the previous versions, aside from increasing the number of stars in the sample, the homogeneity and accuracy of the fundamental $\teff$ and $\logg$ are significantly improved thanks to the high quality of interferometric data and of the Gaia photometric and astrometric data.
This paper follows the spirit of the previous series of papers and presents the chemical analysis of iron-peak and $\alpha$-capture elements: Fe I, Fe II, Sc II, V I, Cr I, Mn I, Co I, Ni I, Mg I, Si I, Ca I, Ti I, and Ti II.
To do so, we performed systematic queries in different available public archives of high-resolution ($R>42,000$) optical spectrographs complemented by our own observations. 
We compiled a total of 2,539 individual spectra, which resulted in 522 combined spectra from the different instruments of the full sample of 192 GBS v3, including the Sun, and for ten additional stars that were part of the previous GBS versions but are not considered GBS because their angular diameters have an indirect measurement.
We analysed these spectra using different methods to provide the community with a complete spectral library and detailed abundances for these reference stars.

This paper is organised as follows.
We describe the source of the observational data and the pre-processing of the data to obtain a homogeneous spectral library in Sect.~\ref{sec:data} and the spectroscopic analysis in Sect.~\ref{sec:handling}. 
In Sect.~\ref{sec:Solar} we provide details on the line selection used to derive the mean elemental abundances and the results obtained for the solar spectra.
In Sect.~\ref{sec:metallicities} we describe the resulting Fe abundances of the full sample of stars, and Sect.~\ref{sec:abundances} presents the results of the individual element abundances.
In Sect.~\ref{sec:errors} we analyse the different sources of uncertainty in our abundances.
Finally, Sect.~\ref{sec:data_availability} contains the details of the accessibility of the data, and conclusions are described in Sect.~\ref{sec:conclusions}.

\section{Observational data}\label{sec:data}

The GBS are nearby and bright stars. Thus, in general, they have been observed extensively with different high-resolution spectrographs.
That is why the main source of our observational material are spectra retrieved from public archives.
We have queried the archives of the following high-resolution optical spectrographs ($R>42,000$): HARPS, UVES, and FEROS using the ESO phase3 spectroscopic specialised interface\footnote{\url{https://archive.eso.org/wdb/wdb/adp/phase3_spectral/form}}; NARVAL and ESPaDOnS using the PolarBase database\footnote{\url{http://polarbase.irap.omp.eu/}} \citep{Petit+2014}; and the ELODIE archive via its public webpage\footnote{\url{http://atlas.obs-hp.fr/elodie/}} \citep{Moultaka+2004}.
From these public archives, we obtained 2,335 spectra corresponding to 182 stars.

We also performed our own observational runs to complement this sample, with the UVES, FIES, HERMES and CAFE spectrographs.
We particularly observed the remaining 20 stars, but we also re-observed several stars for comparison purposes.
In total, we have obtained 204 spectra corresponding to 51 different stars.
These observations allowed us to complete the observations of the full GBSv3 sample of 201 stars presented in \citet{Soubiran+2024}.
We add to our sample of spectra the high-S/N solar spectra used in the first version of the GBS (Paper II): four from HARPS, one from NARVAL and one from UVES.

\subsection{Instruments and data reduction}

HARPS \citep{Mayor+2003} is an ESO facility mounted at the 3.6m telescope in La Silla, and it is mainly dedicated to the measurement of radial velocities for exoplanet search.
The spectral range covered is 378-691 nm with a resolving power of 115,000.

UVES \citep{Dekker+2000} is the high-resolution optical spectrograph of the Very Large Telescope (VLT), located at its second unit.
We have used the 580 setup which gives as a result the spectrum divided in two parts covering from 476 to 580 nm (lower part) and from 582 to 683 nm (upper part).
The resolving power of the instrument depends on the slit used, going from $\sim$40,000 to $\sim$100,000.

FEROS \citep{Kaufer+1999} is a spectrograph installed at the MPG/ESO 2.2-metre telescope located at ESO’s La Silla Observatory.
It covers the full optical spectrum from the near-ultraviolet to the near-infrared $\sim$350-920 nm with a resolving power of around 48,000.

Reduced, calibrated, extracted and merged 1D (unnormalised) spectra from HARPS, UVES and FEROS are provided in the ESO spectroscopic archive, obtained with the ESO science grade pipelines.
We additionally obtained a spectrum with a dedicated DDT program (ID 112.26WR) for HIP98624, in this case we used the public ESO UVES v6.4.1 pipeline to obtain the 1D spectrum.

NARVAL \citep{Auriere2003} and ESPaDOnS \citep{Donati+2006} are two twin spectro-polarimeters mounted at the 2.03 m Telescope Bernard Lyot (TBL) at the Pic du Midi Observatory, and at the 3.6m Canada France Hawaii Telescope (CFHT) located atop the summit of Mauna Kea.
They both are capable of obtaining a full optical spectrum from 370 to 1,050 nm with a resolving power between 68,000 and 80,000, depending on the observation mode.
Spectra from both instruments are automatically reduced, calibrated and extracted with dedicated data reduction pipelines based on the Libre-ESpRIT software \citep{Donati+1997}.
We have used our own pipeline, based on iSpec, to merge the spectra orders by performing a weighted mean in the overlapping regions.

ELODIE \citep{Baranne+1996} was an echelle spectrograph with resolution of $\sim$42,000 used at the 1.93 m telescope of Observatoire de Haute Provence (OHP) between late 1993 and mid 2006. The 1D spectra provided in the ELODIE dedicated archive \citep{Moultaka+2004} are resampled in wavelength with
a constant step of 0.005 nm covering the range 400–680 nm and
given in “instrumental” flux, homogeneously treated by a procedure developed by \cite{Prugniel+2001}.

FIES \citep{Telting+2014} is a high-resolution echelle spectrograph mounted at the 2.5 Nordic Optical Telescope (NOT) at the Roque de los Muchachos Observatory (ORM), with a maximum spectral resolution of $\sim$67,000 covering the entire spectral range 370-830 nm.
The reduction, extraction calibration and order merging of the obtained spectra was done automatically in the telescope with the FIESTool software.

HERMES \citep{Raskin+2011} is a highly efficient echelle spectrograph, which became the workhorse instrument of the 1.2 m Mercator telescope on the ORM, covering the entire wavelength range from 380 to 900 nm with a spectral resolution of 85,000.
The obtained spectra were automatically reduced, extracted, calibrated and order-merged with the HERMESDRS pipeline.

CAFE \citep{Aceituno+2013} is a high-resolution (60,000) spectrograph covering the visible range (400-920 nm), mounted at the 2.2 m telescope of the Centro Astronómico Hispano en Andalucía (CAHA) or Calar Alto Observatory.
The spectra were reduced with the CAFE data reduction pipeline \citep{LilloBox+2020}, which reduces, extracts, calibrates and merges the orders obtained with the spectrograph.

\subsection{Combined spectra}
\begin{figure}[htp]
\centerline{\includegraphics[width=0.4\textwidth]{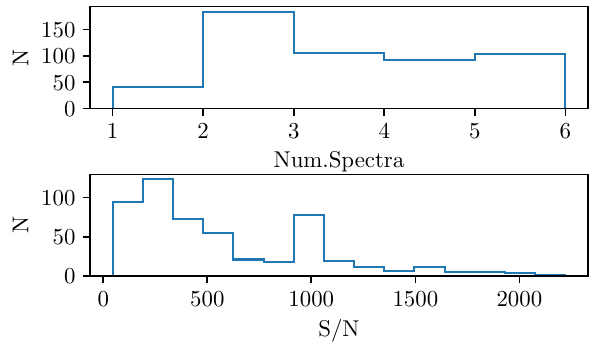}}
\caption{Distribution of the number of combined spectra per instrument and star (top), and S/N distribution of these combined spectra (bottom).}
\label{fig:Nspec_SN}
\end{figure}

We queried the aforementioned archives, selecting individual spectra with a minimum signal-to-noise ratio (S/N; coming from the data reduction pipelines) of 20, and we co-add all the un-normalised spectra from the same instrument (see Sect.~\ref{sec:preproc}) to reach a minimum S/N$\sim$100.
We obtained a total of 522 combined spectra for the 202 stars (including the Sun).
Figure~\ref{fig:Nspec_SN} shows the distribution of the number of spectra from the same instrument that were used for each combined spectrum and the S/N distribution of all the combined spectra.
Most of the stars are observed with more than one instrument (not shown in the figure), and only 30 stars have a single observation.
In particular, three stars (HIP 12114, HIP 57757 and HIP 79672) have been observed with six instruments.

Most of the combined spectra have a S/N well above 100, with the mean S/N being $\sim$600 and the maximum being $\sim$2,200 for HIP 37826 (Pollux).
By construction of the query of spectra in the public archives, there is a peak of the S/N distribution at 1,000.
This is because there are several stars for which we found a particularly large number of spectra of the same instrument available, generally coming from exoplanet search projects.
To avoid downloading and combining too many spectra, we only selected the best S/N ones up to when we could reach a combined S/N of 1,000.

\section{Data handling and analysis}\label{sec:handling}
We analysed the full sample of spectra using an automated pipeline based on iSpec \citep[version 2020.10.01;][]{BlancoCuaresma+2014,BlancoCuaresma2019}.
iSpec is a Python software designed to perform operations on stellar spectra and to compute radial velocities, atmospheric parameters, and chemical abundances using different atmospheric models and radiative transfer codes.
The used pipeline consists of a first pre-processing of the data to ensure a homogeneous treatment of the spectra from different instruments and a subsequent common strategy for the spectroscopic analysis.

\subsection{Pre-processing}\label{sec:preproc}
The spectra were obtained with nine different instruments, each one with different technical specifications and different conventions (e.g. the presence of an explicit error spectra, and specific wavelength handling).
This is why, to construct a spectral library and to perform a homogeneous spectroscopic analysis, we used a common pipeline developed in \citet{BlancoCuaresma2018} and later adapted in \citet{Casamiquela+2020}, which performs a pre-processing of the spectra described as follows:

\begin{itemize}
    \item Each spectrum is cut to a restricted common wavelength range (480 to 680 nm) and downgraded to a common resolution (42,000). A homogeneous resampling of 0.001 nm steps is also performed.
    \item Negative fluxes are set to zero, and errors are estimated in a common manner using the S/N with the dedicated native iSpec function (which essentially divides the flux by the S/N).
    \item Telluric regions are identified by a cross-correlating with a telluric line list and masked, and the fluxes are set to zero.
    \item For the particular case of UVES spectra, the red- and blue-arm spectra for the same star are sometimes provided in separate files; in these cases, the spectra are merged into one, inserting a gap (of zero flux) between them.
    \item The different spectra from the same star and instrument are co-added to increment the final S/N. To do so, one spectrum is selected as reference, then we perform a cross-correlation of all other spectra of the same star against the reference. Finally, we sum up the absolute fluxes of all aligned spectra.
    \item Radial velocity determination is performed with the co-added spectra from cross-correlation with a high S/N solar spectrum from NARVAL, as described in \citet{BlancoCuaresma+2014}. The resulting radial velocities are used to correct the co-added spectra to laboratory wavelengths, ready for the spectroscopic analysis. Heliocentric radial velocities are derived from this procedure after applying the barycentric correction. However, for spectra obtained with HARPS and ELODIE, we adopt the mean of the individual radial velocities available from public archives, as these values are determined by dedicated velocimetric pipelines expected to provide the highest precision.
    \item The continuum is determined homogeneously, computing a synthetic spectrum using the fundamental $\teff$ and $\logg$ derived by \citep{Soubiran+2024}. The observed and synthetic spectra are divided, and the result is smoothed with a median filter to determine the continuum placement. Finally, the spectra are then normalised with the determined continuum using the native iSpec function.
\end{itemize}

Our pipeline is not optimised to analyse spectroscopic binaries, but our algorithm to derive the radial velocity (which uses the iSpec native function \emph{cross\_correlate\_with\_template}) is able to automatically detect double-line spectroscopic binaries.
Once the velocity profile was constructed from the cross-correlation process, local minima were searched in the smoothed profile (of twice the resolution).
The mean velocity was calculated by fitting a Gaussian profile near the main peak.
If other peaks are found with a probability above 50\%, other Gaussian profiles are fitted.
We have found a star in the GBS v3 sample which shows a secondary peak in the cross-correlation function (CCF): HIP 109176.
It is an F dwarf with roughly solar metallicity for which we have a NARVAL spectrum (with very high S/N: 1683) and 38 HARPS spectra of various S/N ranging from 60 to 230.
The CCF of the NARVAL spectrum and of most of the individual HARPS spectra show a faint secondary peak, indicating that two sets of lines at different radial velocities are visible.
When we visually inspected the two co-added spectra, we saw that they differ significantly, even though we could identify common lines from the main component.
This is possibly due to the different phases in which the double-line spectroscopic binary was observed and to the fact that the HARPS spectrum results from the co-addition of 38 spectra taken in different phases.
We flag the abundance results of this star in our final tables.

There are another 16 stars in the GBSv3 sample identified as spectroscopic binaries in the literature: HIP 171, HIP 27913, HIP 37279, HIP 3850, HIP 4151, HIP 45343, HIP 46853, HIP 5336, HIP 5445, HIP 61317, HIP 67927, HIP 71683, HIP 72567, HIP 81693, HIP 86614, HIP 92512.
We did not see any relevant feature in the CCFs nor the spectra of these stars, indicating that the contribution of the companion is possibly too faint to be detected.

\subsection{Spectroscopic analysis}\label{sec:analysis}

We employed the spectral synthesis method to compute the atmospheric parameters and individual chemical abundances using four different available radiative transfer codes: SPECTRUM \citep[version 2.77;][]{Gray+1994}, Spectroscopy Made Easy \citep[SME; version 574;][]{Valenti+1996,2017A&A...597A..16P}, MOOG \citep[version November 2019;][]{Sneden+2012}, and Turbospectrum (version 19.1; \citealt{Alvarez+1998,Plez2012}; see also \citealt{Gerber+2023}).
All of the codes assume local thermodynamic equilibrium (LTE) and solve the equation of radiative transfer within a one-dimensional hydrostatic stellar atmosphere. Atmospheric models with plane-parallel geometry are used for stars with a surface gravity of $\logg$=3.0 or higher, while spherical models are used at lower $\logg$ values. In the latter case, the radiative transfer calculations are done in spherically symmetric geometry for two of the codes (SME and Turbospectrum), and in plane-parallel geometry for the others (SPECTRUM and MOOG).
Further differences between the codes are in part of technical\footnote{SPECTRUM and SME are written in the C programming language, while Turbospectrum and MOOG are written in Fortran.} and numerical nature.
For example, the algorithms used for solving the equation of radiative transfer range from an attenuation operator method with a Bézier spline approximation for the source function in the case of SME to a Feautrier scheme including scattering in the case of Turbospectrum.
Further aspects of line formation which may be treated differently in each code concern the calculation of the continuous opacity and the computation of molecular equilibrium.
A subtle difference between MOOG and the others, already mentioned and discussed in \citealt{Jofre+2017} and \citealt{BlancoCuaresma2019}, is that it does not recompute electron densities, but uses the values provided by the input model atmosphere.
Concerning the line opacity, differences may arise from the ``computation radius'' used for each line\footnote{This implies a decision on which transitions to include at each wavelength point, see e.g. \url{https://www.appstate.edu/~grayro/spectrum/spectrum276/node23.html}.}, and the calculation of the line profile including treatment of radiative and collisional broadening.
Finally, the calculation of the surface flux from specific intensities may be implemented in different ways, e.g. by integration over the stellar disk or applying a limb darkening correction.
For details the reader is directed to the publications referred to for each code and \citealt{BlancoCuaresma2019}, as an in-depth description would go beyond the scope of this article.

For all codes, we use the MARCS\footnote{http://marcs.astro.uu.se/} atmospheric models \citep{Gustafsson2008}, the solar abundances by \citet{Grevesse+2007}, and the sixth version of the line list from the \emph{Gaia}-ESO survey \citep{Heiter+2021}.
In the case of Turbospectrum, in addition to the atomic lines, we use the Gaia-ESO molecular line lists as compiled by Thomas Masseron (private communication), which are also publicly accessible\footnote{\url{https://keeper.mpdl.mpg.de/d/6eaecbf95b88448f98a4/}} \citep{Gerber+2023}.
This molecular information represents more than 3 million transitions, in contrast to the $\sim$140,000 atomic transitions in the \emph{Gaia}-ESO atomic line list. Turbospectrum is specifically optimized to handle such a vast number of transitions efficiently. The other radiative transfer codes, however, are not designed for molecular line lists of this magnitude, and their inclusion would lead to computationally prohibitive processing times. Therefore, the use of these molecular data was restricted to the Turbospectrum analysis.

For all four codes, spectral fitting was performed by comparing observed fluxes (weighted by their uncertainties) to synthetic spectra of the star's $\teff$ and $\logg$, computed at the observed resolution (42\,000) for pre-selected line regions.
Atmospheric parameters and chemical abundances were varied in two separate steps until convergence was reached using a least-squares algorithm.
The fitting is performed only in a set of pre-selected line regions provided by the iSpec default installation, selected to give accurate chemical abundances for the Sun.
Two line selections are provided for each code: one optimized for determining atmospheric parameters and a second to be used for a line-by-line determination of individual chemical abundances \citep[see][for further details]{BlancoCuaresma2019}.
For each line region, a cross-correlation with the synthetic template is performed to allow small line shifts, and afterwards, the line mask is automatically adjusted to avoid nearby overlapping lines.
A basic quality control of the linemasks is done, for instance, rejecting lines detected at too high line shifts (there is a risk of line misidentification), lines for which the flux at the detected peak is null, linemasks which overlap with the peak of a telluric line, lines with too poor fits (high root mean square) or for which the peak of the flux is too different with respect to the peak in the synthetic spectrum.

As for the atmospheric parameters, $\teff$ and $\logg$ were fixed to the fundamental values derived by \citet{Soubiran+2024}, and each code infers independently $\mh$, $\alph$, as well as the microturbulence parameter $\vmic$.
The overall metallicity fitted in this step is inferred from the atmosphere models: a zeropoint to the solar mixture, which is obtained by fitting at the same time all linemasks.
Since we do not only use Fe lines in this procedure, the $\mh$ is not necessarily equal to the $\feh$, even though they are similar.
The $\alph$ abundance obtained in this step depends on the previous overall metallicities, following the predefined $\alph$ enhancement at low metallicities from the MARCS models.
In this work, we do not consider these $\mh$ and $\alph$ values, since we obtain the elemental abundances in a second step instead.
Concerning the broadening effects, the projected equatorial rotational velocity $\varv\sin i$, the macroturbulence parameter, and the spectral resolution are degenerated and difficult to disentangle.
We applied a similar strategy as in \citet{BlancoCuaresma2019}: we used a fixed value for $\varv\sin i$ of 1.6 \kms and the macroturbulence (fixed to 0 \kms), and only the spectral resolution at which the synthetic spectra are generated was let free, accounting for all broadening effects.
For the Sun, we used the fixed atmospheric parameters $\teff=5771$ K and $\logg=4.44$ dex, based on the GBSv1 recommended values of \citet{Heiter+2015}.

The absolute chemical abundances of individual lines are measured in a second step using the atmospheric parameters fixed to the values resulting from the previous step.
To ensure reliable measurements, we only consider lines whose reduced equivalent widths (REWs\footnote{$REW=\log_{10}(\mathrm{EW}/\lambda)$ where EW is the equivalent width and $\lambda$ is the wavelength}) fall within the range $-6.7<\mathrm{REW}<-4.5$.
This criterion helps exclude lines that are either too weak, which can be affected by noise, or too strong, which can suffer from saturation effects.
The choice of these limits is not straightforward in a sample with such a variety of stellar parameters and metallicities, but it is a trade-off between the number of lines detected and the accuracy of the resulting abundances.
In particular, the lower cut in REW is motivated by the fact that we are able to measure very weak lines for high S/N spectra, which is particularly crucial for metal-poor stars.
However, it tends to select very weak lines in moderately noisy spectra, which do not give precise abundances (particularly for solar and metal-rich stars).
Thus, to not compromise accuracy when the S/N is not that high, we have added an additional cut in the depth of the line with respect to the continuum, compared with the overall S/N: $\mathrm{depth}>3\sigma$, where ($\sigma = \frac{1}{\mathrm{S/N}}$).

After the previous filtering, for each spectrum, we use the line selection detailed in Sect.~\ref{sec:linesel}, and we apply a $2.5\sigma$ clipping to discard lines which may be affected by cosmics or other issues in particular spectra.
This is only done for spectra and elements which have more than 5 lines measured.
Finally, we compute the average absolute abundance per spectrum and element, and we use the solar abundances for each code derived in Sect.~\ref{sec:Solar} to compute solar-scaled [X/H] values.
We also compute abundances per star (for each code) obtained by averaging the values obtained by the different spectra and the scatter among the spectra.

\section{Solar abundances and line selection}\label{sec:Solar}

In this section, we detail the analysis of the line-by-line chemical abundances obtained for the six solar spectra from UVES, HARPS, and NARVAL using the different radiative transfer codes.
We explain the criteria we used for the line selection, and we later compare the abundance results with literature.

\subsection{Line selection}\label{sec:linesel}
We used the solar spectra to perform a selection of lines per element and code based on the line coherence given by the six solar spectra.
We based our selection on the flags provided by the Gaia-ESO survey line list study \citep{Heiter+2021}.
For each studied line, a \emph{loggflag} and a \emph{synflag} were provided, with possible values: `yes' (Y), `undetermined' (U), and `no' (N).
These flags indicate the level of accuracy of the atomic $\log gf$ value and the level of blending from neighbouring lines, respectively.
We excluded from our selection all lines with \emph{loggflag}=N for having an unreliable determination of their $\log gf$.

In this work, we focus on the analysis of iron-peak and $\alpha$ elements, which have a minimum of one line flagged as \emph{loggflag}=Y and \emph{synflag}=Y (YY).
We based the selection of lines using as reference the YY lines.
We then proceeded by keeping additional lines that give abundances similar to the reference lines (up to $\pm0.05$ dex) and that give similar dispersions as the reference lines among the six spectra (up to $2\sigma$).
We highlight the following particular cases:
\begin{itemize}
    \item The two YY lines for \ion{Si}{I} (569.0425 and 570.1104 nm) are very often affected by telluric regions in our full sample of spectra. Thus, for homogeneity reasons, we discard those lines, and instead, we keep the seven lines which are compatible with absolute abundances from the literature for the Sun (see next subsection) and compatible among them. They all have \emph{loggflag}=Y.
    
    \item The two YY lines for \ion{Sc}{II} (565.7896, 566.7149 nm) give consistent abundances among them but 0.15 dex larger than literature values. We were unable to determine the cause of this discrepancy; consequently, we decided to use the YU/YN lines 523.9813, 552.679, 566.9042, and 568.4202 nm, which give very coherent abundances and are consistent with literature values.
    
    \item For \ion{Mg}{I} the two YY lines correspond to the Mg triplet, which are too strong and out of the REW range considered in this work. There are five other Mg lines measured: on one side, the 550.9597 and 578.5313 do not have any defined flag and are both in a crowded region; and on the other side, we find the 571.1088 (UY), 631.8717 (UU), 631.9237 (UY) with defined flags. 631.9237 is slightly blended, and both 631.9237 and 631.8717 provide very high absolute abundances with respect to the literature for all four radiative transfer codes, around 7.75 compared to usual literature values of 7.45-7.5 \citep[e.g.][]{Magg+2022}. This issue seems to come from a small offset between the continuum of the synthesised spectra and the observed spectra in the small region [631.6,632.2] nm. Since the continuum of the synthetic spectra is slightly higher than the observed one, a larger abundance is needed to provide a good fit at the centre of the line. This effect is particularly relevant for dwarfs because these two lines are much weaker than for the giants. We have checked that performing the fit with a higher resolution (HARPS spectra at 115,000) does not solve the problem and that this difference in the continuum is not instrument-dependent. Additionally, this region is in the middle of strong telluric bands. Thus, we speculate that this continuum difference may be due to telluric contamination or a lack of information on the underlying line list. Since iSpec's best results are provided for 571.1088, we decided to keep only this line. Still, we warn the reader that this line gives compatible solar abundances for the literature only for TURBOSPECTRUM (see Fig.~\ref{fig:overall_literature}) and an overestimation of the Mg abundance for the other three codes. We highlight that the abundances obtained from the Mg triplet fit are very close to the literature values, but, as mentioned before, these two lines do not satisfy the established upper limit in REW and thus are discarded from our final results.
\end{itemize}

\subsection{Comparison with literature}

\begin{figure*}[h!]
\centerline{\includegraphics[width=\textwidth]{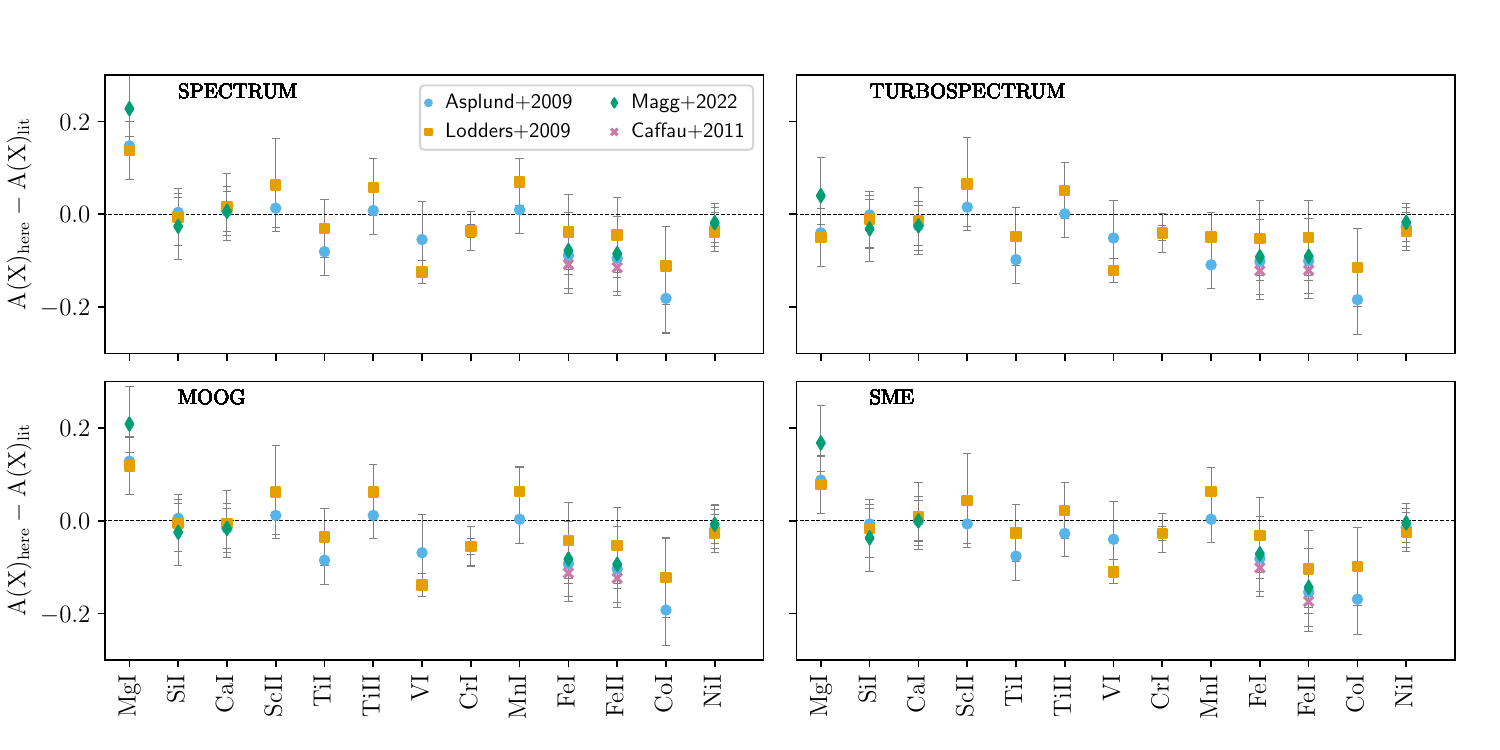}}
\caption{Literature comparison (with different colours and symbols) of the mean solar abundances obtained here for the four radiative transfer codes (in different panels). Each point represents the mean abundances obtained for all solar spectra, averaging all the measured lines. }
\label{fig:overall_literature}
\end{figure*}

We present in Fig.~\ref{fig:comp_lit_1} and ~\ref{fig:comp_lit_2} a line-by-line comparison of the absolute\footnote{Absolute abundances expressed in the form of $\log (N_X/N_H)+12$, with $N_X$ and $N_H$ the number of absorbers of the element X, and of hydrogen, respectively.} abundances obtained in this work for the previously selected lines, with several solar values from literature \citep{Asplund+2009,Caffau+2011,Lodders+2009,Magg+2022}.
Fig.~\ref{fig:overall_literature} presents the overall literature comparison of the mean solar abundance values obtained for each radiative transfer code.
Each point represents the average among the six solar spectra of the mean line abundances.
The displayed errors in the Y axis are computed as $\sqrt{\delta A_{here}^2 + \delta A_{lit}^2}$, with $\delta A_{here}$ being the error computed from the product of posterior distribution functions as explained in Sect.~\ref{sec:errors_final}, and $A_{lit}$ the errors quoted in the literature.

As seen in Fig.~\ref{fig:overall_literature}, we see an overall good agreement with the literature, with a slight tendency of our solar abundances to be underestimated for all radiative transfer codes (overall average of $-0.04$ dex).
In particular, for \ion{Fe}{I}, we obtain a mean difference of -0.08 with all literature values and all codes (very similar value for \ion{Fe}{II}, except for SME which produces a larger offset), though it is compatible at 1$\sigma$ of the uncertainties.
We highlight the case of Mg for which the selected line is in agreement with the literature only for TURBOSPECTRUM.
SPECTRUM, MOOG and SME obtain an overestimation of the absolute abundances of around 0.10-0.15 dex for this line.
The other Mg lines cited in Sect.~\ref{sec:linesel} produce even larger abundance offsets compared to the literature.

We detail the mean solar abundances obtained with the four codes in Table~\ref{tab:solar_abus}.
Our uncertainties tend to be small, particularly those computed from the dispersion among the different spectra, which are a reflection of the high S/N of all solar spectra.
The uncertainties coming from the line dispersions are more realistic but are not available for Mg, which only has one single line analysed.
Finally, the mean of the quoted errors is always larger than the dispersion among spectral lines (see discussion in Sect.~\ref{sec:errors_linefits} for a thorough explanation).

\section{Fe abundances}\label{sec:metallicities}

In this section, we study the Fe abundance results per spectrum for all the stars using the mean of the \ion{Fe}{I} abundances and subtracting the solar Fe abundance obtained for each radiative transfer code.
We compare our results with those adopted by \cite{Soubiran+2024}, which are a compilation from the literature mainly based on the PASTEL catalogue \citep{Soubiran+2016}.

\begin{figure}[htp]
\centerline{\includegraphics[width=0.5\textwidth]{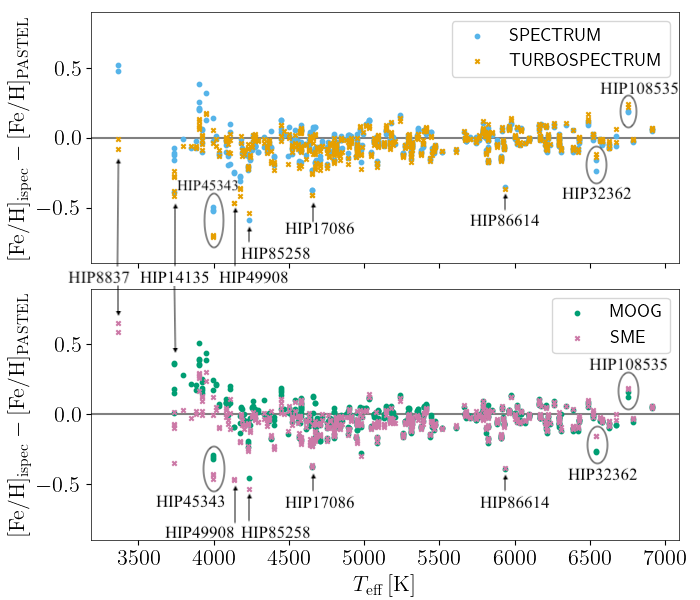}}
\caption{Abundance of Fe per spectrum obtained in this study compared with the one from the literature adopted for the determination of fundamental $\teff$ and logg in \citet{Soubiran+2024} as a function of $\teff$ (fundamental value). The two panels and the different colours show the values from the different radiative transfer codes. Some outliers discussed in the text are indicated (the MOOG value of HIP8837 is outside of the Y-axis limits at $\sim1.3$).}
\label{fig:met_PASTEL}
\end{figure}

We plot the comparison of the Fe abundance as a function of effective temperature in Fig.~\ref{fig:met_PASTEL}, for each radiative transfer code.
We observe a general good consistency between the literature and the values that we have obtained in this work with the four radiative transfer codes.
The mean offsets with respect to the literature are slightly negative, of around -0.04 dex.

A few interesting trends can be seen in Fig.~\ref{fig:met_PASTEL}, particularly for the coolest stars.
We obtained an increasingly large dispersion in the comparison for stars cooler than $\teff\sim5000$ K.
In the case of SPECTRUM, MOOG, and SME, a clear trend towards higher $\feh$ compared to PASTEL is visible at $\teff<4500$ K.
This is possibly due to the increasing importance of undetected or poorly modelled blends in cool stars and, in particular, the presence of molecules.
Consequently, the behaviour of TURBOSPECTRUM, which we use for synthesising molecular lines, is significantly different to the other three codes and shows no overall trend.
However, we obtain a negative zero point of -0.1 dex for stars at $\teff<4500$ K for TURBOSPECTRUM, which for SPECTRUM and SME is -0.05 and -0.03 respectively, and for MOOG is positive (0.08 dex)
A similar zero point for TURBOSPECTRUM was already found in previous GBS versions \citep{Jofre+2014}, where it was run by the ULB method, which used BACCHUS \citep{Masseron+2016}.

There are a few remarkable cases in the comparison to literature, which we highlight in the plot and comment on below (from left to right):

\begin{itemize}
    \item HIP8837 (psiPhe) is the coolest star in our sample, a metal-poor M giant of 3362 K, for which we have a HARPS and a UVES spectrum (the same as GBSv1). It is a particularly difficult case, and the Fe abundance determinations of SPECTRUM, SME and MOOG\footnote{the point of the MOOG comparison for this star is out of the plot limits in Fig.~\ref{fig:met_PASTEL}, the difference is around +1.2 dex} seem highly overestimated. Additionally, the differences obtained between the two instruments are also notable, even though they are compatible within the quoted uncertainties, which are always $>0.5$ dex. Interestingly, the results from TURBOSPECTRUM seem more consistent, and they differ by only 0.1 dex with the only determination of $\feh$ reported for that star in PASTEL, which is the GBSv1 value by \cite{Jofre+2014}. 

    \item HIP14135 (alfCet) is the second coolest giant in the sample (3738 K) for which five spectra are available. The absolute discrepancies with the PASTEL values are overall smaller than in the previous case, though the scatter among instruments is large for SPECTRUM, MOOG and SME (0.12-0.15 dex). Again, TURBOSPECTRUM seems to give a more consistent Fe abundance for this star with 0.07 dex scatter among the five spectra. We note that the nine literature values reported in PASTEL also show a large dispersion of Teff and Fe abundance, with [Fe/H] ranging from -0.60 to +0.02 dex, which reflects the difficulty in obtaining robust abundances for cool stars. 

    \item HIP45343 is the coolest dwarf in the sample (3997 K), for which we have three spectra that give overall consistent abundances. Depending on the code, we obtain Fe abundance between [Fe/H]$\sim-0.15$ and $-0.4$, with TURBOSPECTRUM giving the largest discrepancy with the two literature values from PASTEL, which are $\feh\sim0.17$ and [Fe/H]$\sim0.34$.

    \item HIP49908 is the second coolest dwarf in the sample (4132 K), for which PASTEL gives a mean Fe abundance of $+0.21\pm0.06$ dex. We have spectra from NARVAL and ESPADONS, for which SPECTRUM and MOOG obtain $\feh$ at around 0, and TURBOSPECTRUM and SME around -0.2, in all cases with uncertainties of 0.25 dex.
    
    \item HIP85258 (betAra) is a red supergiant at 4232 K for which we have a HARPS spectrum, the same as GBSv1. For all codes, we obtain similar [Fe/H] of around $\sim0$ with relatively large uncertainties $\sim0.15$, which is consistent with the value obtained in GBSv1 and contrasts with the only other literature value in PASTEL of $+0.5\pm0.1$ dex  (with $\teff=4582$ K, 300 K higher than our value).
    
    \item HIP17086 is a red giant at 4652 K for which we have HARPS and HERMES spectra, and for which all codes give a consistent difference of -0.35 dex with the single literature value listed in PASTEL (with $\teff=4905$ K, 250 K higher than our value). 
    
    \item HIP86614 is part of a binary star with the two components being resolved by high-resolution spectroscopy. We have observed the A component, a subgiant of 5936 K for which we obtain [Fe/H]=-0.4 (from a single NARVAL spectrum). The five literature values in PASTEL are in good agreement, with a mean of $\feh=+0.01$, but with corresponding $\teff$ about 400 K larger than our fundamental value. 
        
    \item We have two spectra (HERMES and NARVAL) from HIP32362, an F dwarf of 6537 K for which we obtain slightly metal-poor values while the literature quotes a single value of $0.14\pm0.1$ \citep{Yee+2017}. This star has very broad spectral lines because it is a high rotator \citep[66 \kms, ][]{Schroder+2009}, and thus our [Fe/H] uncertainties are also of the order of 0.1, larger than other stars at the same temperature.
    
    \item HIP108535 is also a moderately high rotator at 6754 K \citep[26 \kms, ][]{Schroder+2009}. In this case, the abundances that we obtain with the two spectra (FIES and CAFE) give consistent values of [Fe/H]$\sim$-0.05, which are 0.2 dex more metal-rich than the single literature value from PASTEL with a $\teff$ about 600 K lower than our fundamental value.
\end{itemize}

\subsection*{Dependence on the instrument}

We used the sample of stars that were observed with more than one instrument to check the internal consistency of the Fe abundance and any possible instrumental bias.
For each instrument, we compute the difference between the [Fe/H] obtained for the spectrum of that instrument and the value for the same star observed with other instruments.
In the plot, we only show the TURBOSPECTRUM values; see below.

\begin{figure}[htp]
\centerline{\includegraphics[width=0.5\textwidth]{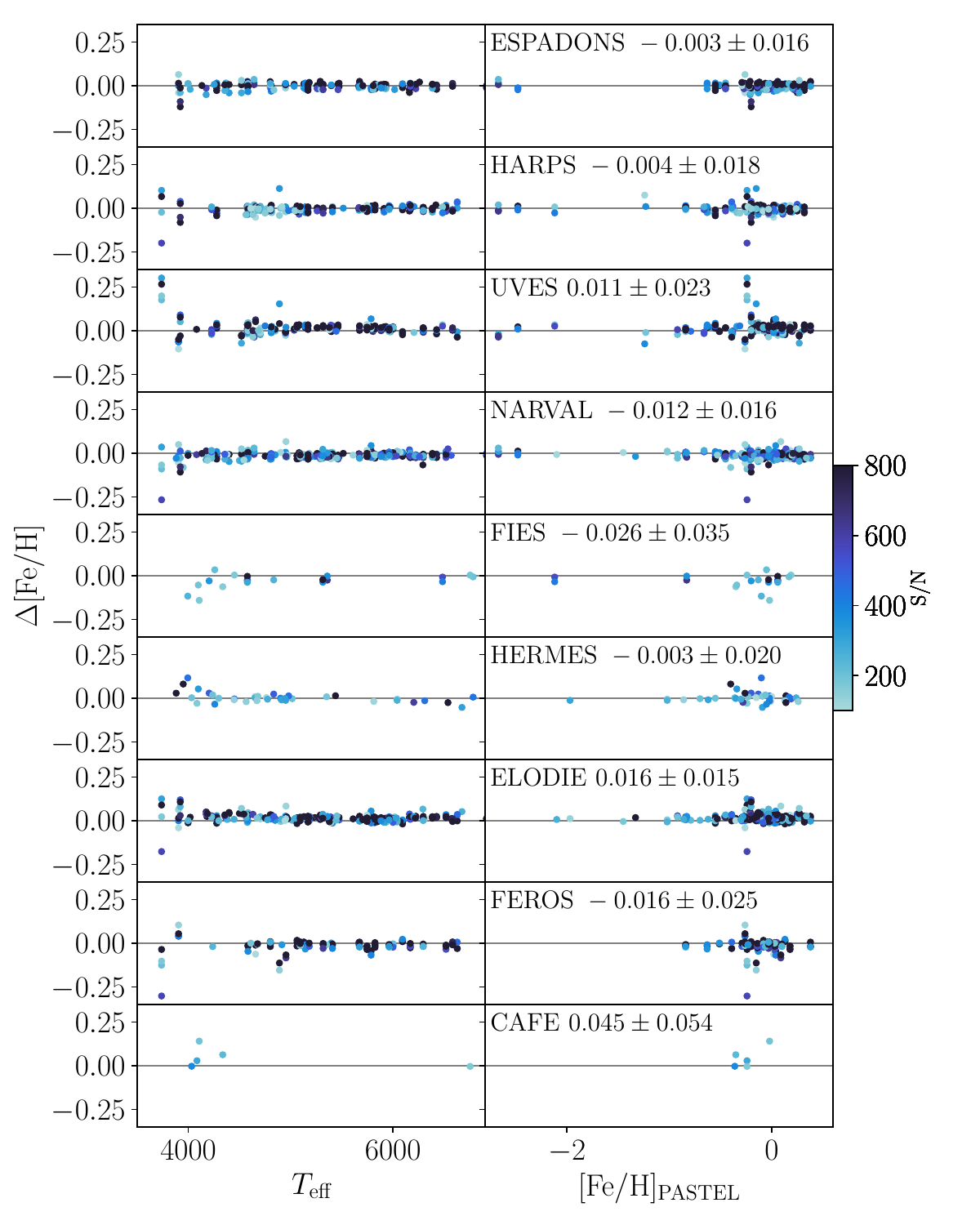}}
\caption{Difference in Fe abundance per instrument (computed as $\mathrm{[Fe/H]_{mean}}-\mathrm{[Fe/H]_{instr}}$) as a function of effective temperature (left) and PASTEL $\feh$ (right). The colour depicts the S/N of the spectra. The average and spreads of the differences for stars above 4000 K are written in each panel.}
\label{fig:comp_instr}
\end{figure}

Figure~\ref{fig:comp_instr} shows the results obtained for the Fe abundance for different instruments.
We see a very high consistency of the Fe abundance in general, better than 0.03 dex.
The plot against effective temperature clearly shows the difficulty in obtaining reliable abundances for $\teff\lesssim 4000$ K, without dependence on the instrument or on the S/N.
We include in Figure~\ref{fig:comp_instr} the average and spread of these Fe abundance differences for stars with $\teff$ above 4000 K.
The mean values are always well below the spreads and, in most cases, of the order of 0.01 dex.
The highest differences are found for FIES and CAFE ($\sim$0.04 dex), probably due to the low number of stars, but, in any case, below the obtained spread.
The right-hand plots in Figure~\ref{fig:comp_instr} show the differences against external Fe abundance (from PASTEL), which do not show any particular bias with [Fe/H].
These results show that there are no significant differences in the Fe abundance obtained with the different instruments, highlighting the quality of the homogenisation process.

We found that the mean differences among instruments shown in Fig.~\ref{fig:comp_instr} are very similar when we use the Fe abundances obtained with the other radiative transfer codes.
However, for stars with $\teff$ below 4000 K, there is a significant improvement in the consistency between instruments when we use TURBOSPECTRUM (see Figure~\ref{fig:comp_instr_cool}).

\section{Element abundances}\label{sec:abundances}

As for iron, we computed the mean abundances for each radiative transfer code for all stars, using the respective solar abundances.
Given the fact that TURBOSPECTRUM solar abundances seem to be more compatible with literature (Sect.~\ref{sec:Solar}), and the Fe abundance of cool stars are also more consistent with literature (Sect.~\ref{sec:metallicities}) as well as more consistent among the different spectra of the same star (Figure~\ref{fig:comp_instr_cool}), we limit the analysis and plots of this section to the abundances produced for TURBOSPECTRUM.
The results for all codes are available at the CDS.

\begin{figure*}[htp]
\centerline{\includegraphics[width=\textwidth]{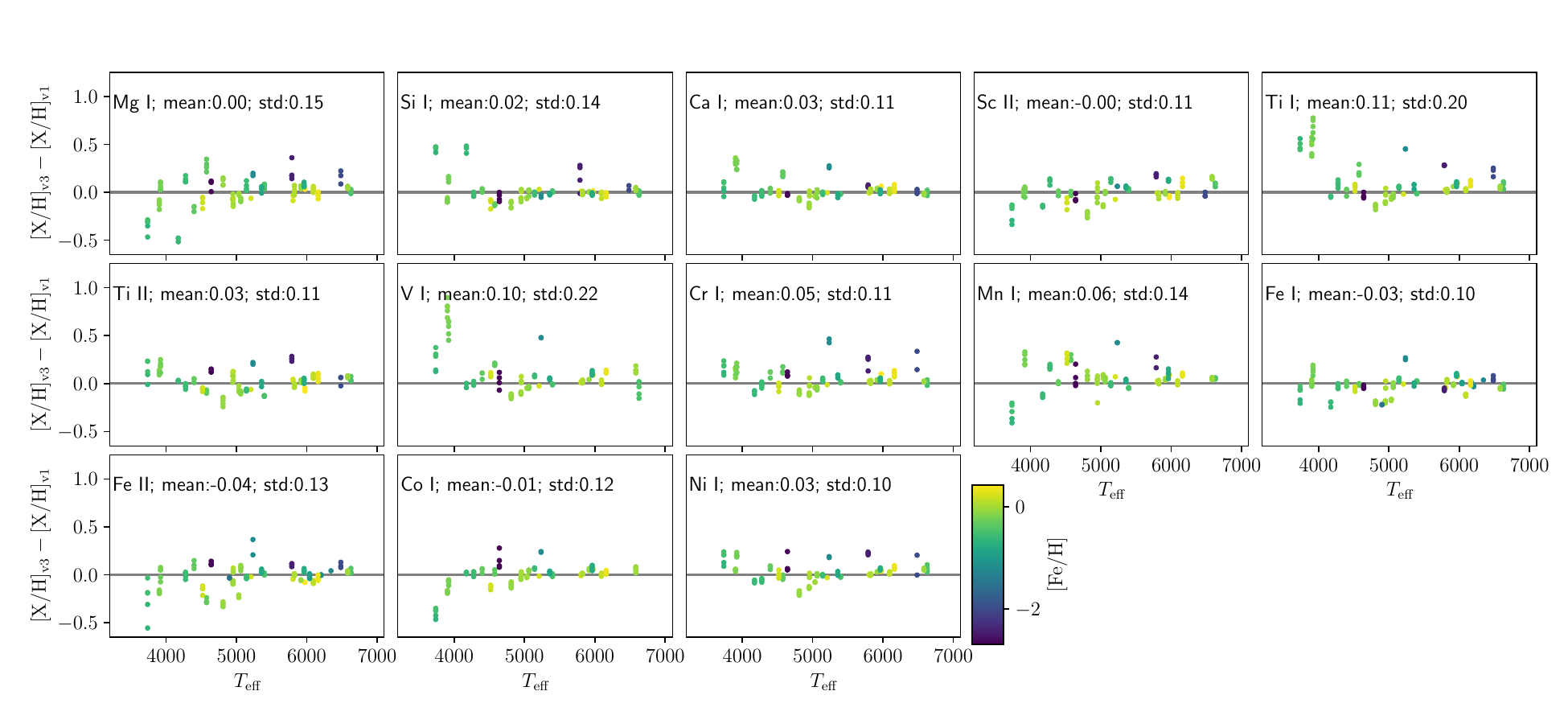}}
\caption{Comparison of the abundances in this study and the values from GBSv1. The mean difference and standard deviation are indicated.}
\label{fig:comp_GBSv21}
\end{figure*}

Figure~\ref{fig:comp_GBSv21} shows the comparison between the abundances derived here, per spectrum, with respect to the values provided in the GBSv1 \citep{Jofre+2015}.
We see a general consistency in the comparison, with the largest offsets ($>0.05$ dex) being for Ti, V and Mn.
These are mainly due to large differences obtained for the three coolest stars (HIP21421, HIP98337 and HIP14135), which also tend to be outliers in other elements (see Mg, Si, Fe II and Co).
This is probably partly because in GBSv1, \citet{Jofre+2015} performed a differential analysis and considered abundances from different radiative transfer codes.
In their fig. 7 \citet{Jofre+2015} shows that there are few offsets between absolute and differential.
Large differences with Ti I compared to Ti II for the three cool stars can also be due to the differential approach that was used in v1 compared to here, or to a different selection of lines which may affect more stars of a certain $\teff$.

We have several stars in common with APOGEE DR17 and GALAH DR4, which are mainly stars that were included in GBSv1.
We obtained an overall agreement between the elemental abundances of these surveys and our values (Figure~\ref{fig:comp_surveys}), particularly with GALAH, where all median offsets are smaller than 0.05 dex except for V and Co (both around 0.14 dex with larger scatter than the other elements).
The persistent outlier in the GALAH comparison is psiPhe, for which GALAH values are much more metal-rich (-0.05 dex) than our determination (-1.3 dex).
Interestingly, this star was found to be more metal-poor in the previous GALAH release (-1.2 dex).
We have more common stars with APOGEE, and the comparison shows larger offsets with respect to GALAH's one, sometimes above 0.1 dex, and a marked trend for Mg and Si towards the metal-poor stars.
This trend is a known issue \citep[see e.g.][]{Soubiran+2022} and seems to be also present in other large surveys, highlighting the difficulty in automated analysis and the importance of having well-characterised metal-poor benchmarks to understand the systematics of the surveys.

\begin{figure*}[htp]
\centerline{\includegraphics[width=\textwidth]{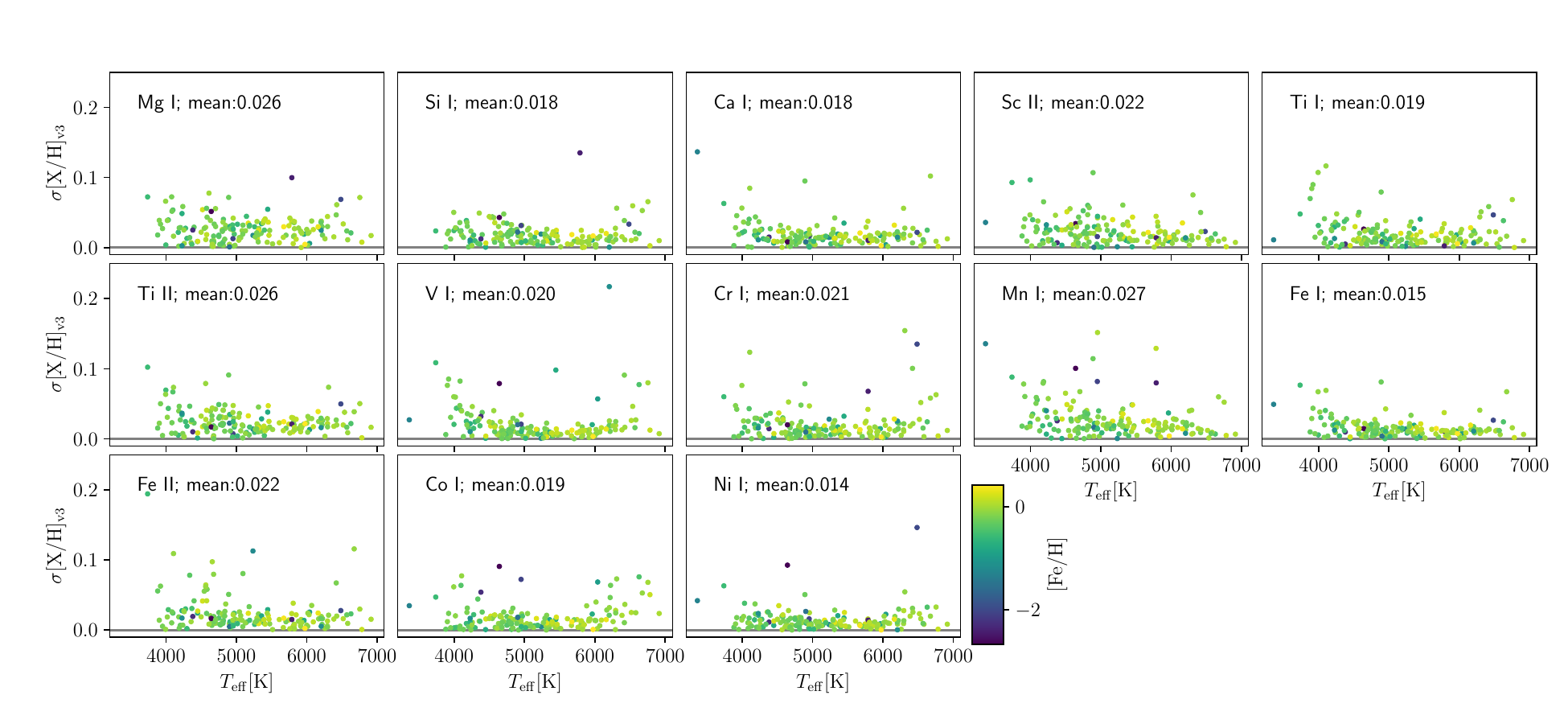}}
\caption{Scatter in abundance obtained among the different spectra of the same star as a function of $\teff$ coloured by the mean [Fe/H] abundance.}
\label{fig:scatter}
\end{figure*}

We also plot in Figure~\ref{fig:scatter} the scatter in abundance obtained among the different spectra of the same star as a function of $\teff$.
We generally obtain very low dispersions with mean values around 0.02 dex, particularly in the region between 5000 and 6000 K.
Mn, Ti II and Mg are the elements with the largest scatters, which is a direct consequence of the lower number of lines detected for them (1 for Mg, and between 3 and 4 for Mn and Ti II, see Figure~\ref{fig:scatter_Nlines}), and which makes the determinations of the abundances intrinsically more uncertain.
The difference in the number of detected lines is also probably the cause of the changes we see as a function of $\teff$, for instance the increase of the scatter for hot stars $\teff>6000$ K (particularly in Ti, V, Cr, Co) and for cool stars $\teff \lesssim 5000-4500$ K.
Very metal-poor stars also tend to give larger scatters for the same reason, and they particularly stand out in the plot for Cr and Ni (see also the three most metal-poor stars in Fig.~\ref{fig:comp_GBSv21}).
The coolest star in the sample HIP8837 (psiPhe) often appears as an extreme outlier in the plots, which is not surprising given its low effective temperature, as we already mentioned in the discussion of Sect.~\ref{sec:metallicities}.

\section{The error budget}\label{sec:errors}
In this section, we aim to analyse the different sources of uncertainty that can affect the abundance values, following the suggestion of \citet{Jofre+2019}.
Primarily, each fit to a line has a quoted abundance computed by iSpec, which depicts the standard error of the line fit (Sect.~\ref{sec:errors_linefits}).
When quantifying the uncertainty associated with the abundance of a given element and star, one can also take into account the abundance dispersion given by the different lines (Sect.~\ref{sec:errors_dispersion}).
This gives a measure of the differences that may arise from the characterisation of each line, which can be due to inaccuracies from the line list parameters (systematic uncertainties) or to the differences in the goodness of fits (random uncertainties).
Finally, the abundances depend to a large extent on the values of the atmospheric parameters adopted.
In this study, we have not determined $\teff$ and $\logg$, but we have assumed their values from the fundamental parameters determined independently from spectroscopy.
These values have associated uncertainties, which we propagate to assess their contribution to the global uncertainty budget in Sect.~\ref{sec:errors_AP}.
In the following subsections, we describe these three sources of uncertainty, and we describe the computation of final error values that we quote in Sect.~\ref{sec:errors_final}.

\subsection{Errors coming from the line fits}\label{sec:errors_linefits}
The uncertainties associated with individual spectral line fits in iSpec originate from the least-squares minimisation process employed to derive the best-fitting abundance.
The individual line abundance error stems from the covariance matrix estimated during the fitting procedure using the Levenberg–Marquardt (LM) algorithm, implemented in MPFIT\footnote{Originally, the routine (named LMFIT) was part of the MINPACK-1 FORTRAN package. It was converted to IDL by Craig B. Markwardt, translated to Python by Mark Rivers, and later optimized by Sergey Koposov (see \url{https://github.com/segasai/astrolibpy/tree/master/mpfit}).}.

For each spectral line, the code models the observed flux by fitting a synthetic spectrum generated using a given set of stellar parameters.
The fitting process minimizes the residuals between the observed and synthetic fluxes by adjusting the abundance of the chemical element under consideration.
The uncertainty of the line abundance is derived from the standard error of the fit, which is computed as follows:

\begin{equation}
\delta_j = \sqrt{C_{jj}},
\end{equation}

where $C_{jj}$ is the diagonal element of the covariance matrix corresponding to the abundance parameter of the given spectral line.
This covariance matrix was approximated as the inverse of the Hessian matrix, $\mathbf{H}$, which in turn is estimated from the Jacobian matrix, $\mathbf{J}$:

\begin{equation}\label{eq:hessian_approximation}
\mathbf{H} \approx \mathbf{J}^T \mathbf{J}.
\end{equation}

However, these error estimates often appear overestimated when compared to the standard deviation of abundances derived from multiple lines of the same element.
This discrepancy arises due to several contributing factors:

\begin{itemize}
\item Correlated noise: The LM algorithm assumes that the residuals are independent and normally distributed, but in practice, spectral noise can be correlated (e.g. after resampling, resolution degradation via convolution, continuum normalisation), leading to an overestimation of uncertainties.
\item Numerical approximation in the covariance matrix: The covariance matrix is estimated numerically from finite differences, which can introduce small systematic deviations from the true error structure.
\item Non-linearity effects: The covariance matrix is evaluated at a single point (the best-fit parameters), whereas the true function may exhibit significant curvature beyond this local region, leading to nonlinearity effects.
\end{itemize}

Consequently, while the LM algorithm provides a statistically sound method for estimating parameter uncertainties via the covariance matrix derived from the Jacobian, simplifying assumptions and systematic errors in the data may lead to overestimated uncertainties for an individual line.
A more accurate estimation of uncertainties would require a full Bayesian treatment or Monte Carlo simulations, both of which are more computationally expensive.
Alternatively, when possible, a more realistic assessment of abundance uncertainties for the mean abundances of an element is often obtained by computing the standard deviation of abundances derived from multiple lines of the same element, which inherently includes both random and systematic effects (see next subsection).
Understanding these discrepancies is crucial when comparing the uncertainty of a single line fit to the overall abundance precision obtained from multiple spectral lines.

\begin{figure}[htp]
\centerline{\includegraphics[width=0.5\textwidth]{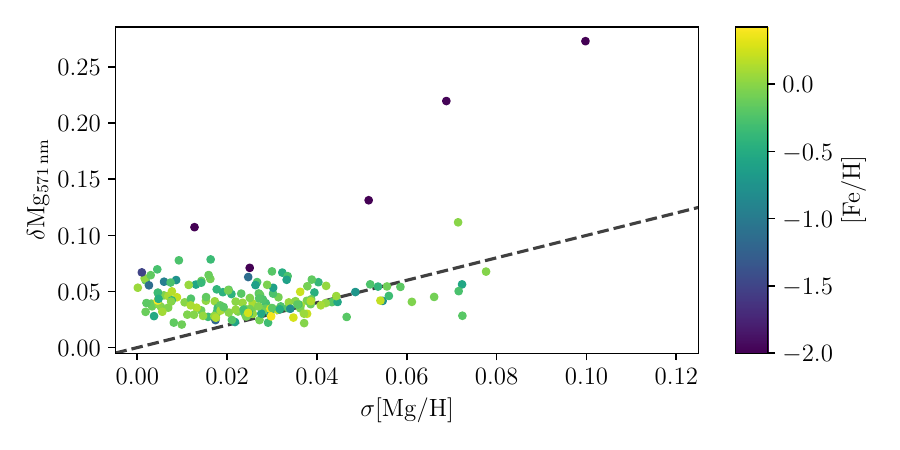}}
\caption{Mean of the quoted error of the 571 nm Mg line abundance (values from Turbospectrum) on the y-axis for stars with more than one spectrum, compared to the scatter in the derived Mg abundances among the different spectra on the x-axis, colour-coded by the mean Fe abundance derived in this study. The dashed line depicts the 1:1 relation.}
\label{fig:unc_Mg}
\end{figure}

In the case of only being able to measure one single line of a given element (like Mg in the case of this study, where we only use the 571 nm line), the dispersion of the abundances derived from different spectra provides a measurement of at least the random uncertainties affecting the line fit.
We plot in Figure~\ref{fig:unc_Mg} the distribution of the quoted error to the Mg line fit ($\delta$Mg$_{\rm 571nm}$), compared to the scatter in the Mg abundance obtained for stars with more than one spectrum ($\sigma$[Mg/H]).
A general overestimation is seen with respect to the obtained scatters, particularly when scatters are very small (the errors seem to saturate at 0.05 dex) or for metal-poor stars (probably because the lines are particularly weak compared to the quoted flux noise).
However, for a given spectrum and element, the errors coming from the line fits are comparable and correlate very well with the depth of the lines.
Thus, we consider that it is safe to use them as an internal relative scale of the quality of the line fit for a given spectrum (see Sect.~\ref{sec:errors_final}).

\subsection{Errors coming from the line-to-line dispersion}\label{sec:errors_dispersion}
\begin{figure}[htp]
\centerline{\includegraphics[width=0.5\textwidth]{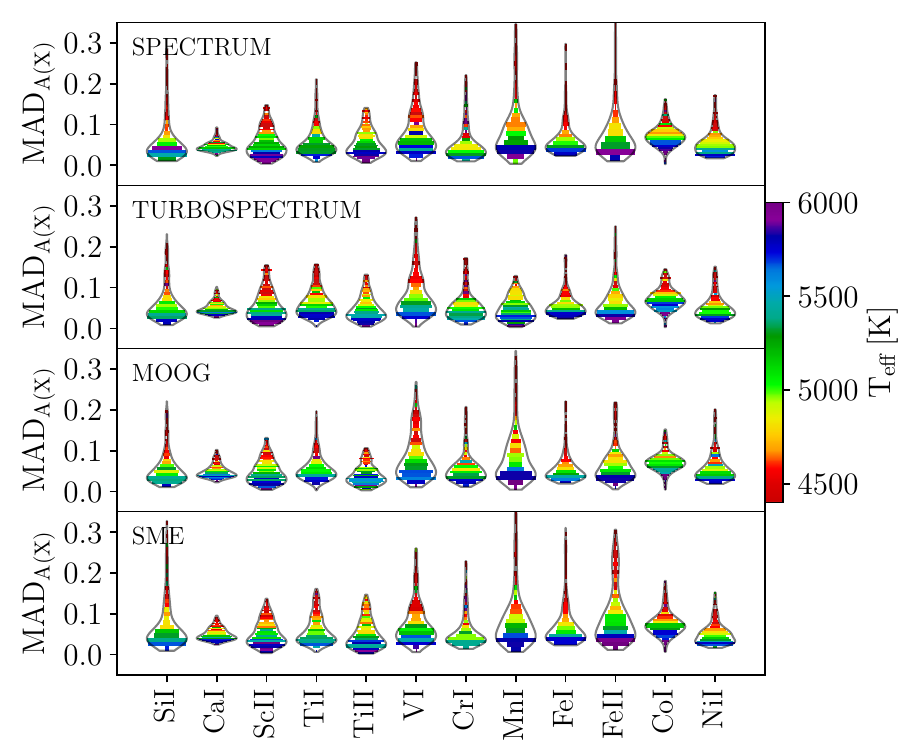}}
\caption{Violin plot representing the kernel density estimation of the distribution of the line scatters (computed as the MAD) per element and radiative transfer code. To visualise the dependency of temperature, we overplot corresponding histograms, coloured-coded by the mean $\teff$ in each bin.}
\label{fig:distrib_unc}
\end{figure}

We plot in Figure~\ref{fig:distrib_unc} the distribution of the mean absolute deviation (MAD) per spectrum and element in the form of violin plots.
This is a direct measurement of the coherence of the different selected lines for a given element.
A vertical histogram is overplotted, showing the dependence in $\teff$ of the obtained MAD.
Mg is not plotted because only one line is available.

The obtained distributions tend to peak at around 0.05 dex and then show more or less prominent tails towards larger dispersions.
V, Mn, Co and Fe II are the elements for which the median of the distributions are larger (around 0.06 dex) for all four codes.

There is a clear dependence of the obtained MAD on the $\teff$ of the star, with cool stars $\teff<5000$ K always giving the largest dispersions.
The coolest stars ($\teff<4500$ K) shape long tails in the distribution of certain elements (e.g. Fe II, Mn, V, Si), which can reach values of more than 0.1 dex.
As discussed in previous sections, this is somewhat expected due to the intrinsically larger random uncertainties implied in the abundance measurement of cool stars.
On the contrary, solar-type and hotter stars seem to provide the most consistent abundances among the different lines.

\subsection{Errors coming from the uncertainties in the atmospheric parameters}\label{sec:errors_AP}

\begin{table*}[]
    \caption{Sample of stars and spectra used to test the effects of the uncertainties on the atmospheric parameters.}
    \label{tab:stars_mctest}
    \centering
    \begin{tabular}{lrccccccc}
    \hline
    Type     & Star & $\teff$ & $\logg$ & $\feh_{\mathrm{PASTEL}}$       & $\feh_{\mathrm{here}}$& $\sigma_{\mathrm{MC}}\feh$& Instr. & S/N  \\
    \hline
F dwarfs     & HIP67927  & $6161\pm18$ & $3.82\pm0.05$ & $0.25\pm0.01$ & $0.32 \pm 0.04$ & $0.015$ & NARVAL   & 1073 \\
& HIP32851  & $6628\pm89$ & $4.19\pm0.04$ & $-0.39\pm0.03$ & $-0.45 \pm 0.03$ & $0.048$ & HARPS  & 1020\\
G dwarfs     & HIP79672  & $5824\pm30$ & $4.42\pm0.01$ & $ 0.03\pm0.01$ & $ 0.06 \pm 0.01$ & $0.020$ & ELODIE & 236 \\
             & HIP76976  & $5788\pm45$ & $3.75\pm0.09$ & $-2.48\pm0.03$ & $-2.42 \pm 0.04$ & $0.042$ & HARPS & 419 \\
K dwarf      & HIP104214 & $4398\pm34$ & $4.63\pm0.01$ & $-0.13\pm0.03$ & $-0.33 \pm 0.14$ & $0.017$ & ELODIE & 330 \\
Subgiant     & HIP95362  & $5268\pm18$ & $3.77\pm0.07$ & $-0.11\pm0.02$ & $-0.17 \pm 0.02$ & $0.011$ & ELODIE & 226 \\
Clump giants & HIP37826  & $4810\pm14$ & $2.55\pm0.03$ & $ 0.02\pm0.03$ & $-0.07 \pm 0.03$ & $0.010$ & ELODIE & 373 \\
             & HIP92167  & $4902\pm30$ & $2.53\pm0.02$ & $-1.45\pm0.03$ & $-1.51 \pm 0.04$ & $0.028$  & NARVAL & 181 \\
Bright giants& HIP69673  & $4277\pm23$ & $1.58\pm0.07$ & $-0.55\pm0.01$ & $-0.55 \pm 0.04$ & $0.006$  & HARPS  & 382 \\
             & HIP68594  & $4642\pm35$ & $1.32\pm0.03$ & $-2.67\pm0.02$ & $-2.68 \pm 0.05$ & $0.043$  & HARPS  & 318 \\
\hline
    \end{tabular}

\tablefoot{$\teff$ and $\logg$ come from fundamental values \citep{Soubiran+2024}, and we also list the [Fe/H] quoted in PASTEL and the one obtained here for that star from TURBOSPECTRUM. We indicate the dispersion in Fe abundance obtained from the 30 random realisations ($\sigma_{\mathrm{MC}}\feh$)}
\end{table*}

\begin{figure*}[htp]
\centerline{\includegraphics[width=\textwidth]{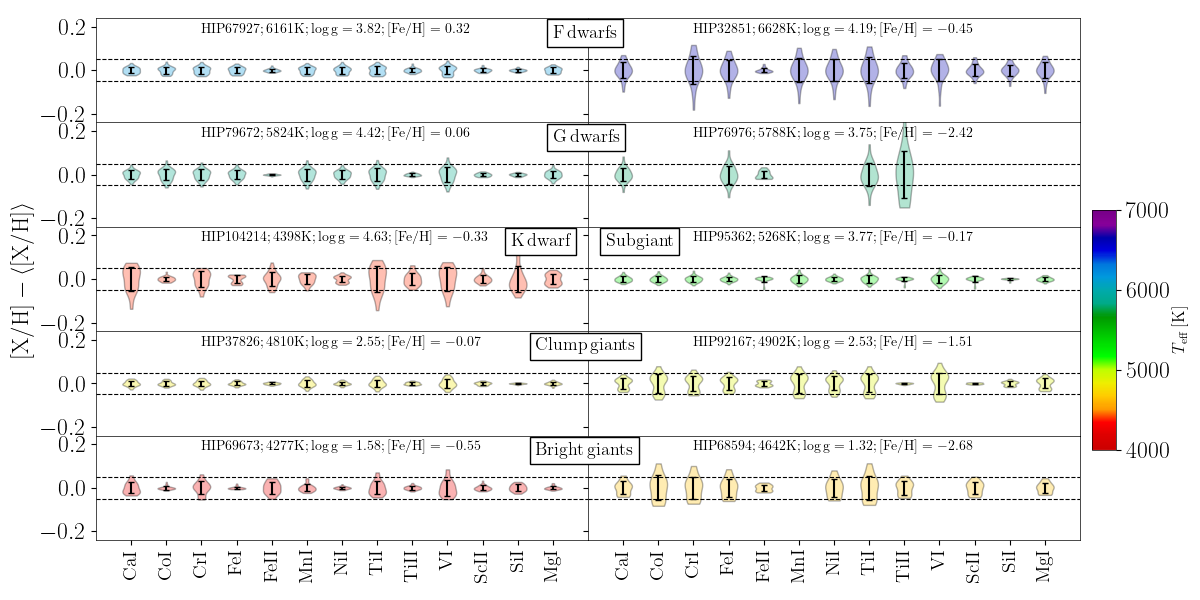}}
\caption{Distributions of the elemental abundances obtained from perturbing the atmospheric parameters within their uncertainties (see Table~\ref{tab:stars_mctest}), coloured by the $\teff$ of the star. The standard deviation for each violin plot is shown as an error bar centred in zero (the mean abundance), and the dashed lines depict the $\pm$0.05 dex level.}
\label{fig:mc_violin}
\end{figure*}

It is well known that the choice of the atmospheric parameters has a large impact on the abundance determination.
The present analysis assumes that each star's effective temperature and surface gravity are known.
These values were obtained independently from the spectra using fundamental methods, i.e. from the limb darkened angular diameter ($\theta_{LD}$) and bolometric flux ($F_{bol}$) to determine the effective temperature, and with the parallax ($\varpi$), angular diameter and mass ($M$) to determine the surface gravity.
This section aims to assess the impact of fixing the $\teff$ and $\logg$ to their fundamental values in the obtained chemical abundances.

Due to the complexity of the interrelations between each quantity and the different fitting procedures used to determine them (see a schematic view in Fig.~\ref{fig:diagram}), the impact of the uncertainties on the chemical abundance values becomes intractable analytically.
However, it can be estimated via Monte Carlo propagation, which does not require knowledge of the correlation matrix.
We thus generated random realisations of $\teff$ and $\logg$ while taking into account the form of the original equations \citep[see sections 3 and 4 from][]{Soubiran+2024}, which make the two atmospheric parameters a function of the following variables:

\begin{equation}
    \teff(\theta_{LD},F_{bol});\,\,
    \logg(\theta_{LD},\varpi,M)
\end{equation}

At the same time, the mass is not an independent variable because it has been obtained with the SPInS code \citep{Lebreton+2020} in the $\teff$-luminosity plane, assuming the $\feh$ from the PASTEL catalogue.
Two resulting masses were obtained using two grids of stellar evolution models, and the mean was taken to compute the fundamental $\log g$.
So, in the end, mass is a function of $M(\theta_{LD}, F_{bol},\varpi,\feh_{\rm PASTEL})$.
Additionally, we point out that $F_{bol}$ also depends on the parallax (through the extinction) and on the $\feh_{\rm PASTEL}$. However, its dependence on the photometry and the used models is much more important for these stars, particularly because their parallax uncertainties are very small, and because the input $\feh$ used has a small impact in the SED fitting \citep[see discussion in sect 3.4 of][]{Soubiran+2024}.
So, for this study, we consider $F_{bol}$ an independent variable.

We thus generated a set of random realisations in the independent parameters ($\theta_{LD}, F_{bol},\varpi,\feh_{\rm PASTEL}$) assuming gaussian uncertainties, and we recomputed $\teff$ and $\logg$ with the same procedure as done in \citet{Soubiran+2024}.
Using this sampling on the atmospheric parameters, we re-did the chemical abundance computation with iSpec.

Due to the high computational cost of running the full pipeline described in Sect.~\ref{sec:handling}, we have limited the random sampling to 30 pairs of $\teff$ and $\logg$, generated for  ten representative stars of the sample.
These stars are listed in Table~\ref{tab:stars_mctest}, and have been chosen to provide good coverage of the $\teff$, $\logg$ and $\feh$ range of the full sample.
Also for computational reasons, we have limited this experiment to one radiative transfer code, TURBOSPECTRUM.

The resulting abundance distributions for each star and element (scaled to the mean element abundance for visualisation purposes) are shown as violin plots in Fig.~\ref{fig:mc_violin}.
The obtained scatter for each star and element is indicated with an error bar, and the scatter value for [Fe/H] ($\sigma\feh_{\mathrm{MC}}$) is listed in Table~\ref{tab:stars_mctest}.
In general, the obtained distributions show small scatters of the order of 0.01-0.03 dex, particularly for cases with solar $\feh$.
The metal-poor stars HIP76976, HIP92167, and HIP68594 show the largest dispersions, highlighting the difficulty of determining their abundance due to the smaller number of available lines.
The largest dispersions are found for HIP32851, a slightly metal-poor F dwarf, but which has the largest uncertainty in $\teff$, resulting in a large range in the input $\teff$ values (6344-6785 K), thus impacting the abundance range.
Remarkably for this star (and also for the other F and G dwarfs), the dispersion obtained of Fe II is much smaller than for the rest of the elements.
This is because of the high sensitivity of the Fe I lines to $\teff$ \citep[see discussion of sec 5.3 in][]{Amarsi+2022}.

Overall, we consider that for the GBS sample, the uncertainties coming from the atmospheric parameters are smaller than the dispersions found among the different lines (see the uncertainty in the [Fe/H] value of the spectrum listed in the table, and discussion in Sect.~\ref{sec:errors_dispersion}).
This is a consequence of the high precision in the $\teff$ and $\logg$ determinations.
Exceptions to this may be the cases highlighted above, particularly when the quoted uncertainties in effective temperature are large ($\sim 80$).
For the GBSv3 sample, the $\teff$ uncertainties range from 5 to 183 K with a median value of 44 K and 15\% of the stars having an error larger than 80 K.

\subsection{Adopted abundances and uncertainties per spectrum and star}\label{sec:errors_final}

We compute, for each spectrum and chemical species, the weighted mean of the absolute abundances for all selected lines, using the inverse of the errors coming from the line fits as weights.
We give as an error of this mean abundance the unbiased weighted standard deviation, computed as 
\begin{equation}
    \sigma^2=\frac{\Sigma w_i (x_i-\mu)^2}{\Sigma w_i - \frac{\Sigma w_i^2}{\Sigma w_i}},
\end{equation}
where $w_i$ are the weights (inverse of the error of each line fit $\nicefrac{1}{\delta_{i}}$), $x_i$ are the line abundances, and $\mu$ is the weighted mean.
This strategy allows for taking into account the goodness of the line fits to compute the mean abundance without overestimating the error budget (see discussion in Sect.~\ref{sec:errors_linefits}).
For the spectra where only one line of the element is measured, the error coming from the line fit is the only uncertainty estimation that we have, even though we warn the reader that it might be difficult to interpret and compare among the different stars.
This is the case of Mg, but particularly for metal-poor stars it can also be the case for other elements for which we only managed to measure one line.
For each spectrum and element, we indicate the number of lines measured.
Finally, the [X/H] values are computed using the solar abundances obtained respectively for each code, and the uncertainties in [X/H] are computed as the quadratic sum of the previously mentioned error and the solar abundance error.

For stars with more than one observed spectrum, we combine the solar-scaled abundances of the different spectra using the product of experts method \citep{Hinton1999}. In other words, we assume that the different measurements of [X/H] follow a Gaussian distribution whose sigma is given by the uncertainty.
Then, we perform the product of the different distributions and we take the mean of the resulting distribution as the final abundance per star, and the standard deviation of the distribution as the uncertainty of the abundance.

\section{Final abundances}\label{sec:final_abus}
\begin{figure*}[htp]
\centerline{\includegraphics[width=\textwidth]{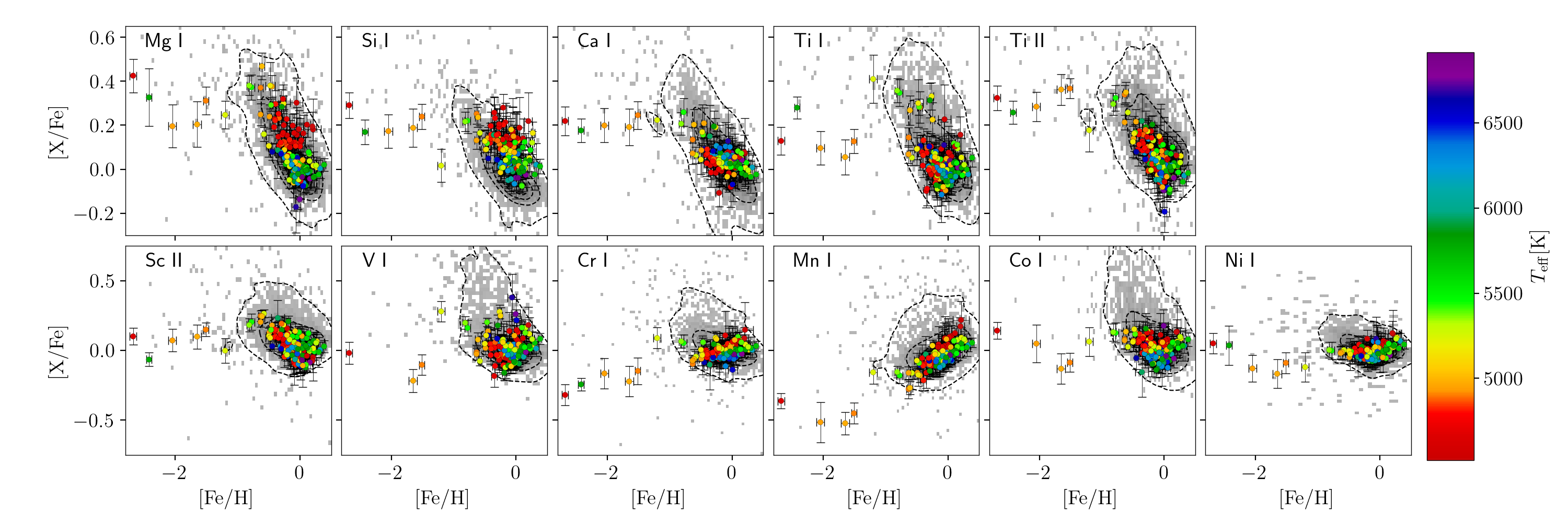}}
\caption{[X/Fe] vs $\feh$ for all the elements analysed in this paper, coloured by effective temperature. The UVES data from the Gaia-ESO survey are overplotted with a 2D histogram and contour lines. Notice the change of the Y scale between the top and bottom panels.}
\label{fig:XvsFe}
\end{figure*}

We explore in Figure~\ref{fig:XvsFe} the [X/Fe] ratios vs [Fe/H] for all the elements we have analysed in this paper.
We only plot here the 192 GBS with direct measurement of the angular diameter \citep{Soubiran+2024}.
To make the plot clearer, we exclude the stars with $\teff\lesssim4500$ K, for which the abundances tend to be less reliable, and the few stars which have very large uncertainties in [X/Fe] ($>0.2$).
We observe similar general behaviours of all the elements when using the other three radiative transfer codes.
We overplot the Gaia-ESO survey stars \citep{Hourihane+2023} (type: "GE\_MW") observed with the UVES setup U580 using a 2D histogram and contour lines.
We have transformed their absolute abundances to solar-scaled using the solar abundances from \citet{Grevesse+2007}.

For all the $\alpha$ elements (top row of Figure~\ref{fig:XvsFe}), we observe a decreasing trend with Fe abundance for the range [Fe/H]$>-0.1$ and a plateau for metal-poor stars.
This trend is consistent with what has been historically seen with solar neighbourhood samples in the literature \citep{Fuhrmann1998}, and it is also seen for the Gaia-ESO stars.
It is produced by the progressive enrichment in iron of the interstellar medium due to thermonuclear supernovae and is well reproduced by Galactic chemical evolution models.
The expected [$\alpha$/Fe] vs $\feh$ trend appears clearer for the GBS when we use Ca or Ti II.
This is probably because these two elements are those for which the line-to-line scatters are the smallest and have less dependence on effective temperature (see Figure~\ref{fig:distrib_unc}).
For Si and Ti I, the tails of the line-to-line scatter distributions of the GBS spread towards larger values (particularly for Si) and are populated by moderately cool stars.
If we do a restrictive cut in effective temperature ($>5000$ K), the trends in these two elements are more clear, but we lose a large fraction of the sample.
In some cases, we see a mild dependence on $\teff$, particularly in Mg and Si ($>0.1$ dex amplitude), which is probably due to the line strength dependence on the atmospheric parameters.
This is why, for Galactic archaeology studies, it is important to take into account these types of effects, or use stars of the same spectral type.
Alternatively, this could be mitigated with a differential analysis.
For the Gaia-ESO stars, the increasing $\alpha$ trend seems to extend to larger $\alpha$ values than for the GBS, particularly for Ca, Si and Ti.
This could be due to the particular selection of stars.

Concerning the Fe-peak elements' dependence with $\feh$ (bottom row of Figure~\ref{fig:XvsFe}), we highlight several behaviours.
Sc shows a decreasing tendency starting at -0.5 dex towards solar values, reproduced by the Gaia-ESO stars, similar to what is seen for $\alpha$ elements, and has been discussed in the literature \citep[e.g.][]{Battistini+2015}.
We observe increasing trends of [Mn/Fe] (which is also visible for the Gaia-ESO stars) and [Cr/Fe], which can partly be due to non-LTE (NLTE) effects.
Indeed, the NLTE corrections for Mn can go up to 0.5 dex at [Fe/H]$\sim-2$ \citep{Bergemann+2008}, and around 0.3 dex at [Fe/H]$\sim-2$ for Cr \citep{Bergemann+2010}.
It remains as future work to perform a full 3D NLTE analysis on the GBSv3 spectra to quantify these trends, which still should be expected to exist due to the differences of supernova yields \citep{andrews17, Kobayashi20}.
A tail of high [V/Fe], up to more than 0.7 dex (with large scatter) towards moderately subsolar $\feh$ is seen in the Gaia-ESO stars, which we do not reproduce with the GBS.
Our V values have a large scatter at around solar $\feh$, which can blur the trend, if it is there.
At metallicities smaller than -1, other studies find [V/Fe] close to Solar: e.g. \cite[][]{Battistini+2015} their fig. 6, and see also \citet{Ishigaki+2013}, which would be in agreement with our three metal-poor stars.
For [Co/Fe], Gaia-ESO stars also show a large decreasing trend towards moderately subsolar $\feh$, which we also see with the GBS but at smaller amplitude. 
This smaller amplitude seems to be in agreement with what other works find \citep[][$\sim$0.2 dex at $\feh$=-1 dex]{Battistini+2015}.
They also find a plateau of [Co/Fe]$\sim0.2$ at lower metallicities in their fig. 6.
Because of the scatter and the few GBS stars at $\feh<-1$, it is difficult to say if our values are compatible with such a plateau.
Finally, a flat trend followed by a slightly increasing trend at super-solar metallicity is seen in [Ni/Fe], not evident for the Gaia-ESO stars, but that has been reported in the literature \citep[e.g.][]{Mishenina+2006}.

\begin{figure}[htp]
\centerline{\includegraphics[width=0.5\textwidth]{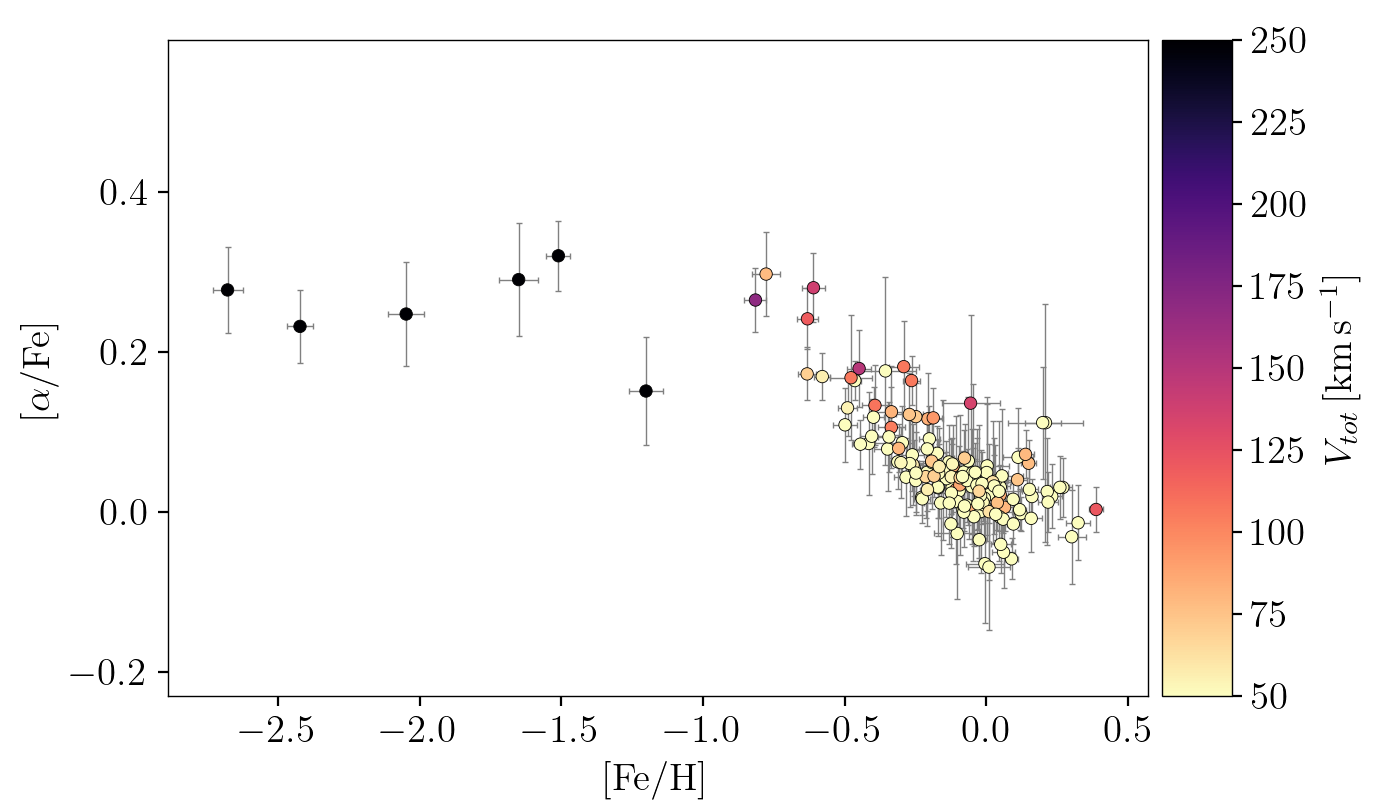}}
\caption{[$\alpha$/Fe] vs $\feh$ coloured by total velocity and with similar restrictions as in Fig.~\ref{fig:XvsFe}, i.e. not-flagged stars with $\teff > 4500$ K and uncertainty on [$\alpha$/Fe] lower than 0.2 dex.}
\label{fig:alpha_fe_vtot}
\end{figure}

In order to illustrate how the GBS sample is representative of the local stellar populations, we show in Fig.~\ref{fig:alpha_fe_vtot} the distribution of [$\alpha$/Fe] vs $\feh$ coloured by the total galactic velocity ($v_{tot}=\sqrt{{\rm U}^2+{\rm V}^2+{\rm W}^2}$).
The global $\alpha$ abundance has been estimated by computing a weighted mean of Si, Ca and Ti (Ti being the weighted mean of Ti I and Ti II).
We have not considered Mg because it is not available for all stars, and it has larger uncertainties.
Galactic velocities (U,V,W) have been computed by combining positions, parallaxes and proper motions from Gaia DR3 \citep{GaiaDR3} (or Hipparcos for the brightest stars) with radial velocities from our spectra.
Figure~\ref{fig:alpha_fe_vtot} clearly shows that the 6 stars with [Fe/H] < -1.0 have kinematics typical of the halo (total velocity $> 300\ {\rm km.s}^{-1}$).
The [$\alpha$/Fe] distribution of these six stars is rather flat with a small dispersion around a mean value of +0.25 dex, consistent with the well-known plateau observed in many spectroscopic investigations of the halo (e.g. \cite{Fuhrmann1998}).
This demonstrates the reliability of our abundances. The rest of the sample extends from about -0.80 to +0.40 in $\feh$ with a decreasing trend of [$\alpha$/Fe] in two parallel sequences.
The upper sequence, with higher $\alpha$ abundances, shows, in general also hotter kinematics characteristic of the thick disc, while the bulk of the sample in the lower alpha sequence shows smaller velocities typical of the thin disc.
There are a few exceptions to the general behaviour of the three populations.
For instance, the most metal-rich star (HIP95447=31 Aql, [Fe/H]=+0.39, [$\alpha$/Fe]=+0.01) has a total velocity of $121 {\rm km.s}^{-1}$ which would make it a good thick disc candidate.
However, the large velocity is only due to its radial component U, which, combined with the super-metallicity with no $\alpha$-enhancement, would suggest that the star was formed in the inner disc and migrated outward in the solar neighbourhood afterwards.
Another notable case is HIP25993, which has thin disc kinematics together with a significant $\alpha$ enhancement [$\alpha$/Fe]=+0.15, but also large abundance uncertainties.
The high $\alpha$ and low $\alpha$ subpopulations in the disc and their correlation with kinematics have been commonly reported in the past from various surveys and high-resolution samples (see, for instance \cite{Soubiran+2005, Bensby+2014, GaiaColRecio-Blanco+2023, Imig+2023}).
Known chemodynamical features are well reproduced by our study and therefore demonstrate that the GBSv3 sample is representative of the stellar populations of the solar neighbourhood.

\section{Conclusions}\label{sec:conclusions}
The third version of the GBS recently released by \citet{Soubiran+2024} provides a step forward in the characterisation of reference stars.
With respect to the previous versions, the GBSv3 sample has improved statistics and better homogeneity and accuracy in the atmospheric parameters $\teff$ and $\logg$.
The present study, the eighth paper of a series, provides a complete spectral library that covers the wavelength range of $480$--$680$ nm and an excellent S/N of 522 normalised spectra for the sample of 192 GBSv3 and for ten additional stars, including the Sun, that were part of the previous GBS versions.
Using the fundamental atmospheric parameters, we performed a chemical analysis of 13 Fe-peak and $\alpha$-capture elements, with the goal of providing to the community reference values for the chemical abundances of these stars.

Our analysis was done with the MARCS model atmospheres \citep{Gustafsson2008} and the sixth version of the \emph{Gaia}-ESO survey line list \citep{Heiter+2021}.
We obtained 1D LTE line-by-line abundances using spectral synthesis fitting with four radiative transfer codes widely used in the literature.
We performed a line selection based on the flags provided with the line list and compared the line abundances obtained from the Sun with literature values.

We discussed how our results for the abundances compare with the literature (Fig.~\ref{fig:met_PASTEL}, \ref{fig:comp_GBSv21}) and showed that, in general, we have small offsets ($<0.05$ dex on average).
We did find a larger scatter and a different behaviour for cool stars ($<4500$ K) with the four different codes.
While SPECTRUM, MOOG, and SME tend to overestimate the Fe abundance for these stars with a marked trend with $\teff$, Turbospectrum seems to give abundances more consistent with the literature and no trends,  possibly due to the inclusion of a molecular linelist.
This is why in the final table we provide the Turbospectrum values as recommended.
The elements that show the largest offsets with the literature are Ti, V, and Mn, and this is in general caused by a large difference obtained for the coolest stars in our sample ($<4000$ K).

We analysed the different sources of error that affect our abundances.
We find that uncertainties due to the errors in atmospheric parameters are, in general, smaller (of the order of 0.01-0.03 dex) than the line-to-line dispersions, which is a consequence of the high precision in the fundamental $\teff$ and $\logg$.
Exceptions to this might be stars with the largest uncertainties in $\teff$.
We observed that the errors quoted for each line fit generally seem overestimated and difficult to compare among the different stars.
However, since they correlate well with the line depth, we used them as weights to compute the mean standard deviation among lines in order to quote an error per element and spectrum.
For the analysed elements, we obtained typical uncertainties of 0.05 dex, and there is a clear dependence of the uncertainty on $\teff$, with cool stars always giving larger values.
Some of these uncertainties can reach values of more than 0.1 dex in certain elements (particularly Fe II, Mn, V, and Si).

We explored the [X/Fe] trends of the GBSv3 in comparison with the Milky Way stars from the Gaia-ESO survey.
In general, we observed a good overlap in the distributions, with some differences that can be explained by the small selection of stars in the GBS compared to the field stars.
Notably, we reproduced the expected [$\alpha$/Fe] versus [Fe/H] trends in the GBS sample for all $\alpha$ elements, with Ca and Ti II being the elements with the clearest trends.
Sc showed a similar trend to the $\alpha$ elements: a decreasing tendency starting at -0.5 dex towards solar values, reproduced by the Gaia-ESO stars and other literature works.
We noted steep trends in [Mn/Fe] (similar to the Gaia-ESO stars) and [Cr/Fe] for metal-poor stars, and we speculate that NLTE effects can be the cause.
V, Co, and Ni showed increasing trends for super-solar metallicity stars, but of much smaller amplitudes.

We also explored the $\alphfe$ versus $\feh$ distribution coupled with the total galactic velocity.
We reproduced the expected behaviour of a local stellar population, with metal-poor $\alpha$-enhanced stars having kinematics typical of the halo and a decreasing trend of $\alphfe$ at $\feh$ between -0.8 and +0.4, in two sequences with a correlation with kinematics that are typical of the thick and thin disks.

The resulting abundances per star, spectrum, and line are available at the CDS.
As future steps, we are going to provide abundances of light and neutron-capture elements, and we plan to do an NLTE analysis, which will complement the set of LTE abundances provided in this work.

\section{Data availability}\label{sec:data_availability}
The data obtained in this paper are only available through the CDS via anonymous ftp to cdsarc.u-strasbg.fr (130.79.128.5) or via \url{http://cdsweb.u-strasbg.fr/cgi-bin/qcat?J/A+A/}.
We publish four tables resulting from our abundance analysis:
\begin{itemize}
    \item The first table contains the final abundances per star for each of the four radiative transfer codes, computed as explained in Sect.~\ref{sec:errors_final}. We provide a column with a flag indicating if the star has an indirect measurement of the angular diameter \citep[not recommended by][]{Soubiran+2024}, or if it is a double-line spectroscopic binary (see discussion in Sect.~\ref{sec:preproc}) and thus also not recommended as a spectroscopic reference.
    \item A second table details the mean abundances per spectrum for each radiative transfer code.
    \item We also make available a table with the absolute abundances and equivalent widths measured for all lines selected in the different spectra and radiative transfer codes, with the references for the gf-values.
    \item Finally, we provide a table with the recommended abundances for each star, the mean $\alpha$ abundances
    and radial velocities measured from our collection of spectra.
\end{itemize}

The first table (per star) and the normalised spectral library at the original resolution of each instrument, and degraded to the common resolution (42,000), are also available on the website \url{https://www.blancocuaresma.com/s/benchmarkstars}.

\begin{acknowledgements}
L.C. and C.S. acknowledge financial support from Observatoire Aquitain des Sciences de l'Univers (OASU). P.J, C.A.G, I.H, S.E and S.V acknowledge support from FONDECYT Regular 1231057. N.L. acknowledges financial support from “Programme National de Physique Stellaire” (PNPS) and from the “Programme National Cosmology et Galaxies (PNCG)” of CNRS/INSU, France. 
A.R.A. acknowledges support from DICYT through grant 062319RA and from ANID through FONDECYT Regular grant No. 1230731.
C.A.G. acknowledges support from  FONDECYT Iniciación 11230741.
U.H. acknowledges support from the Swedish National Space Agency (SNSA/Rymdstyrelsen).
This work was (partially) supported by the Spanish MICIN/AEI/10.13039/501100011033 and by "ERDF A way of making Europe" by the “European Union” through grant PID2021-122842OB-C21; the Institute of Cosmos Sciences University of Barcelona (ICCUB, Unidad de Excelencia ’Mar\'{\i}a de Maeztu’) through grant CEX2019-000918-M; the Horizon Europe HORIZON-CL4-2023-SPACE-01-71 SPACIOUS project funded under Grant Agreement no. 101135205; and the Spanish MCIN/AEI/10.13039/501100011033 through grant RED2022-134612-T.
The preparation of this work has made extensive use of Topcat \citep{Taylor2005}, of the Simbad and VizieR databases at CDS, Strasbourg, France, and of NASA’s Astrophysics Data System Bibliographic Services.
We warmly thank Carme Jordi for her contribution to the observations of the GBS as well as Thomas Nordlander and Clare Worley for the fruitful discussions.
\end{acknowledgements}

\bibliographystyle{aa}
\bibliography{references}

\appendix
\section{Additional figures and tables}\label{sec:appendix}

\begin{table}[htp]
\caption{Solar abundances obtained for the different radiative transfer codes. We provide the final abundance combining the line-by-line weighted mean values of the six analysed spectra of the Sun, and its uncertainties as computed in Sect.~\ref{sec:errors_final}.}
\setlength\tabcolsep{10pt}
\begin{tabular}{l|rr|rr|rr|rr}
 &  \multicolumn{2}{c}{TURBOSPECTRUM} & \multicolumn{2}{c}{SPECTRUM} & \multicolumn{2}{c}{SME} & \multicolumn{2}{c}{MOOG}\\
 &  $A$ &  $\delta$ & $A$ &  $\delta$ & $A$ &  $\delta$ & $A$ &  $\delta$  \\
 \hline
Ca I  & 6.32 & 0.02 & 6.35 & 0.02 & 6.34 & 0.02 & 6.32 & 0.02 \\
Co I  & 4.81 & 0.03 & 4.81 & 0.03 & 4.82 & 0.03 & 4.80 & 0.03 \\
Cr I  & 5.60 & 0.01 & 5.60 & 0.01 & 5.61 & 0.01 & 5.58 & 0.01 \\
Fe I  & 7.40 & 0.01 & 7.41 & 0.01 & 7.42 & 0.01 & 7.41 & 0.01 \\
Fe II & 7.40 & 0.01 & 7.41 & 0.01 & 7.35 & 0.02 & 7.40 & 0.02 \\
Mg I  & 7.49 & 0.01 & 7.68 & 0.02 & 7.62 & 0.02 & 7.66 & 0.02 \\
Mn I  & 5.32 & 0.01 & 5.44 & 0.01 & 5.43 & 0.01 & 5.43 & 0.02 \\
Ni I  & 6.19 & 0.01 & 6.19 & 0.01 & 6.21 & 0.01 & 6.20 & 0.01 \\
Sc II & 3.17 & 0.01 & 3.16 & 0.01 & 3.14 & 0.01 & 3.16 & 0.01 \\
Si I  & 7.51 & 0.01 & 7.51 & 0.01 & 7.50 & 0.01 & 7.52 & 0.01 \\
Ti I  & 4.85 & 0.01 & 4.87 & 0.01 & 4.87 & 0.01 & 4.86 & 0.01 \\
Ti II & 4.95 & 0.01 & 4.96 & 0.01 & 4.92 & 0.01 & 4.96 & 0.01 \\
V I   & 3.88 & 0.02 & 3.88 & 0.02 & 3.89 & 0.02 & 3.86 & 0.01 \\
\hline
\end{tabular}
\label{tab:solar_abus}
\end{table}

\begin{figure*}[htp]
\centerline{\includegraphics[width=\textwidth]{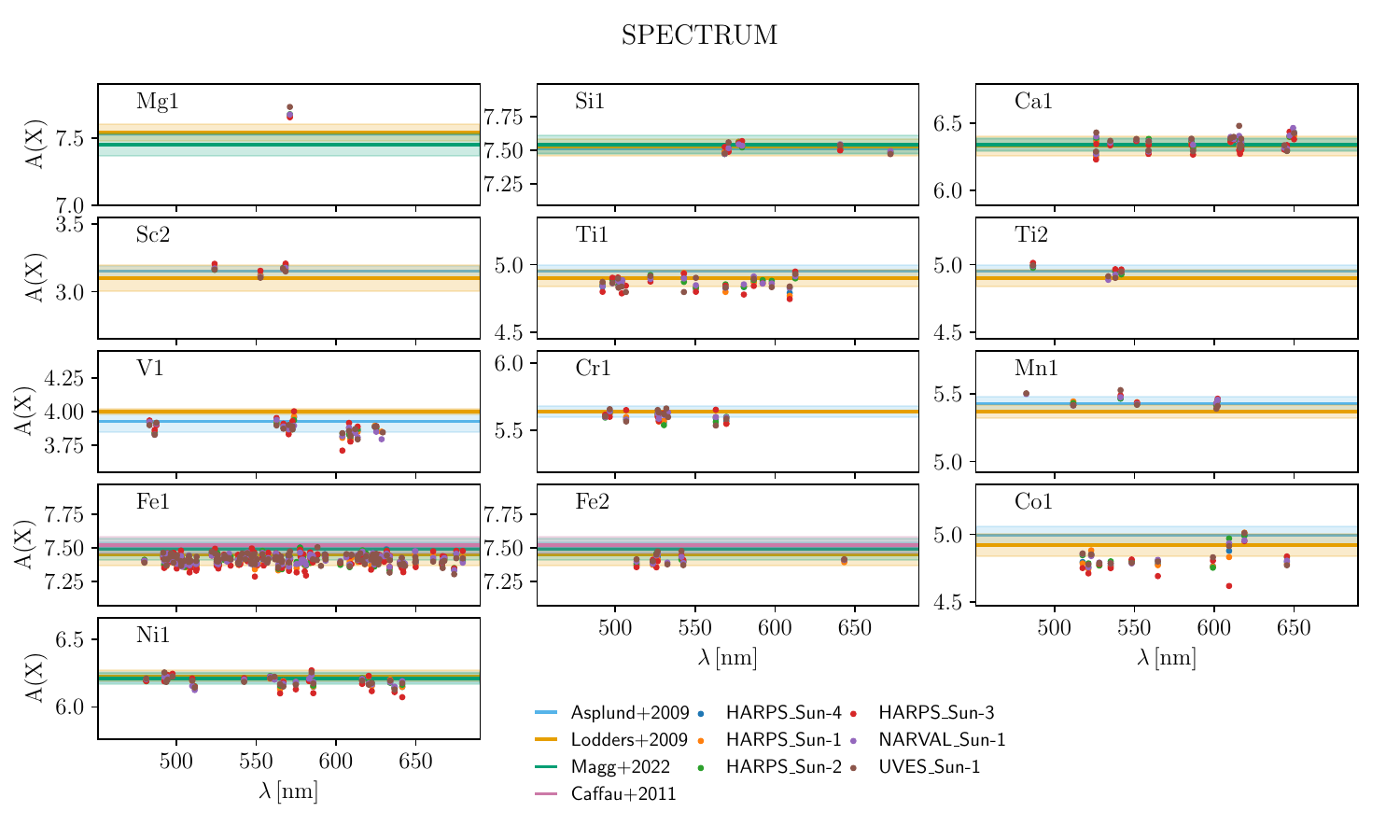}}
\centerline{\includegraphics[width=\textwidth]{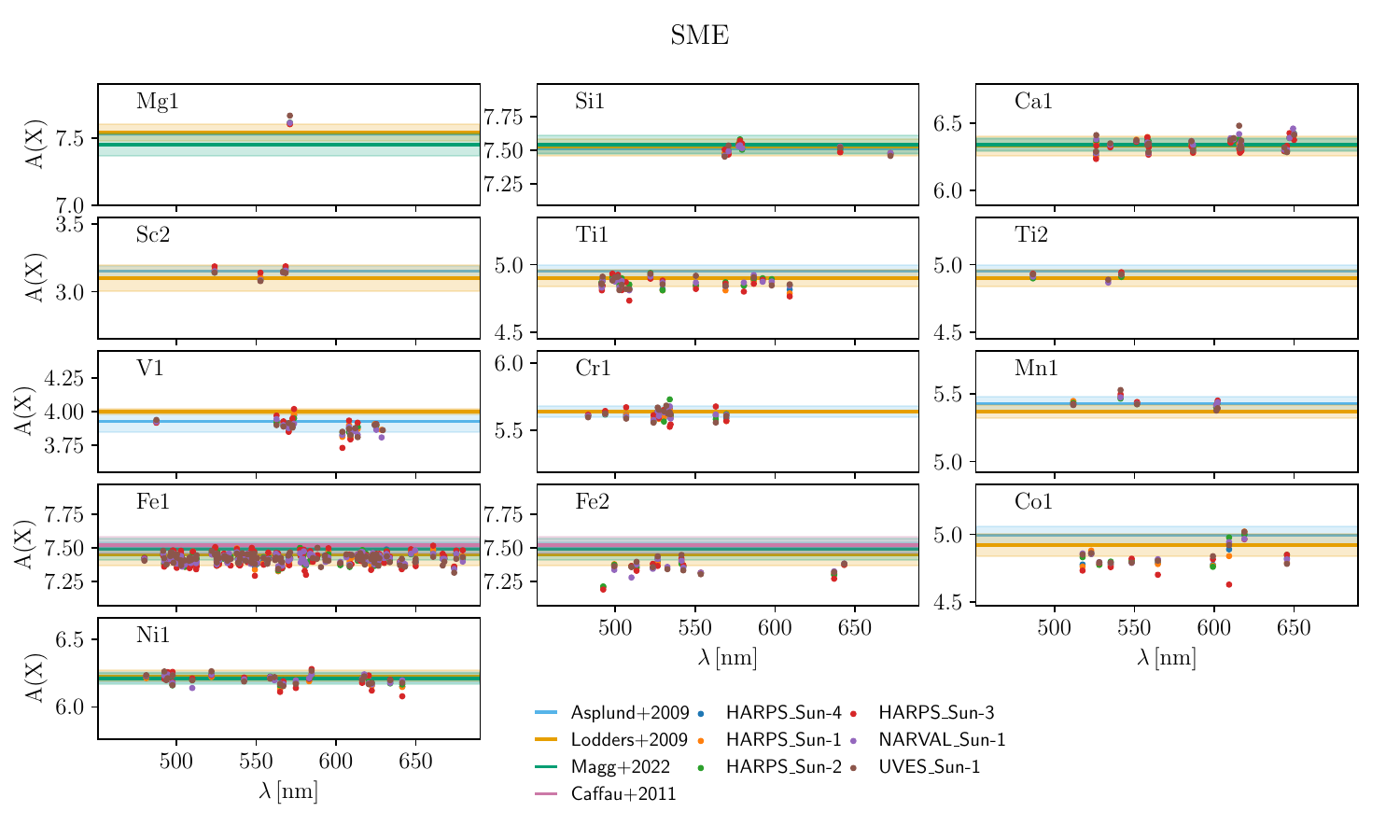}}
\caption{Line by line absolute abundances of the selection of lines performed in Sect.~\ref{sec:Solar} in the six solar spectra. Horizontal lines represent literature values of the solar abundances, with their uncertainties represented as shadowed regions. Selection done for SPECTRUM (top), and SME (bottom)}
\label{fig:comp_lit_1}
\end{figure*}

\begin{figure*}[htp]
\centerline{\includegraphics[width=\textwidth]{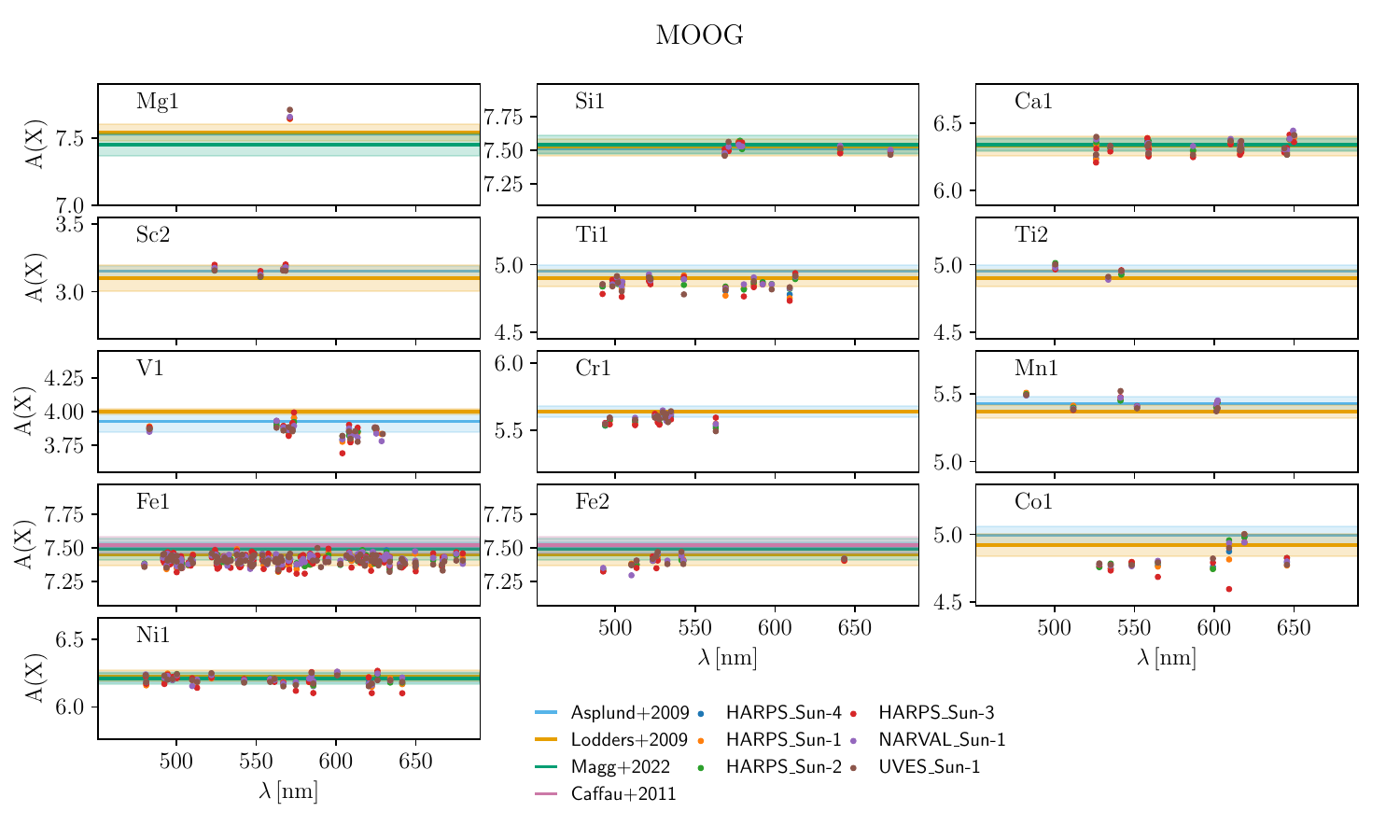}}
\centerline{\includegraphics[width=\textwidth]{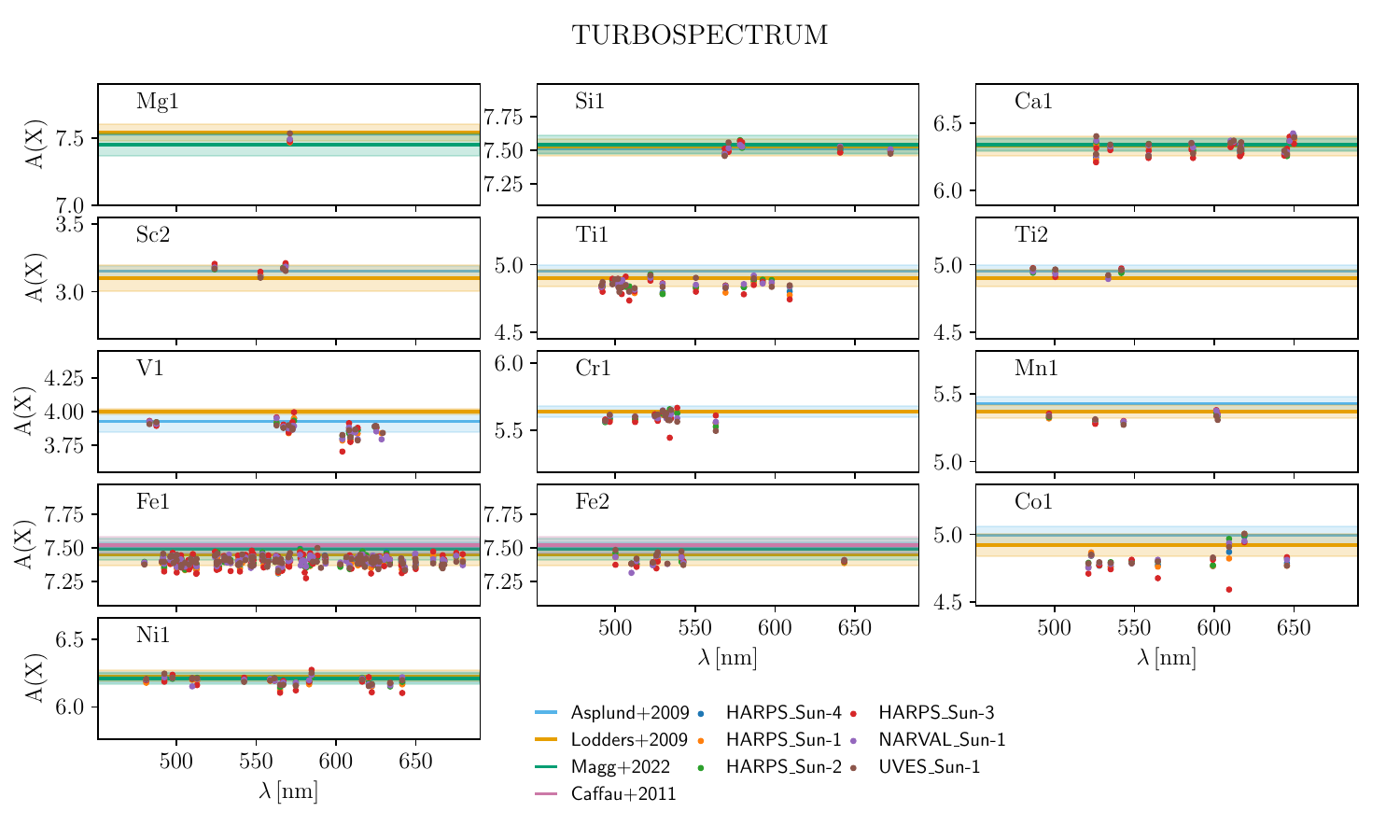}}
\caption{Same as Fig.~\ref{fig:comp_lit_1} for MOOG (top) and TURBOSPECTRUM (bottom).}
\label{fig:comp_lit_2}
\end{figure*}

\begin{figure*}[htp]
\centerline{\includegraphics[width=\textwidth]{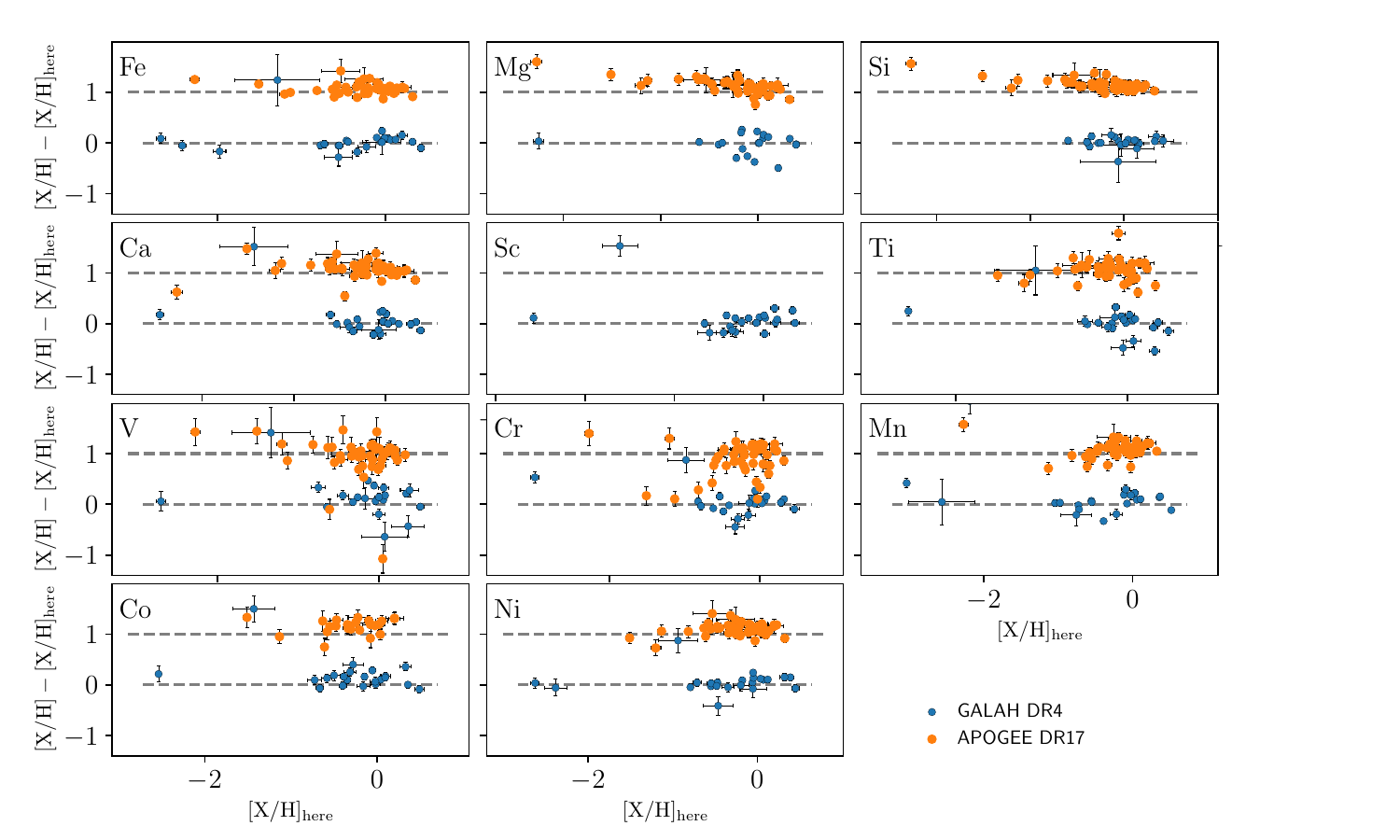}}
\caption{Comparison of the [X/H] values for the common stars between GBSv3 and GALAH DR4 (blue) and APOGEE DR17 (orange). Differences in abundances are plotted against the values from this work. APOGEE's comparison is shifted to 1 for clarity.}
\label{fig:comp_surveys}
\end{figure*}

\begin{figure*}[htp]
\centerline{\includegraphics[width=\textwidth]{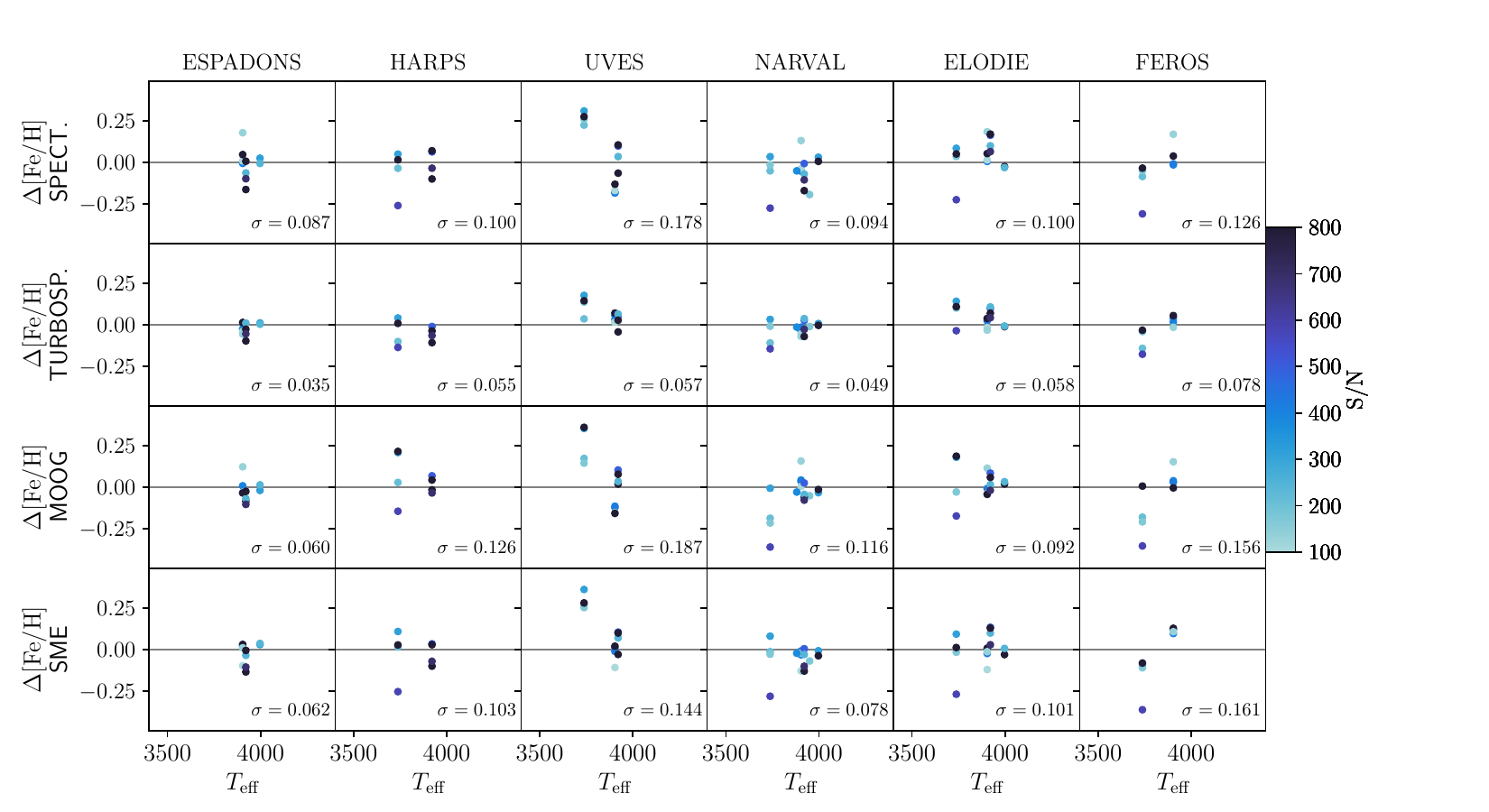}}
\caption{Comparison of the Fe abundances obtained with the different instruments (in vertical panels) and with the different codes (horizontal panels) as a function of effective temperature, only for stars below 4000 K. We only show the instruments which have a significant number of repeated cool stars. The colour depicts the S/N of the spectra. The vertical scatter is indicated in each panel.}
\label{fig:comp_instr_cool}
\end{figure*}

\begin{figure*}[htp]
\centerline{\includegraphics[width=\textwidth]{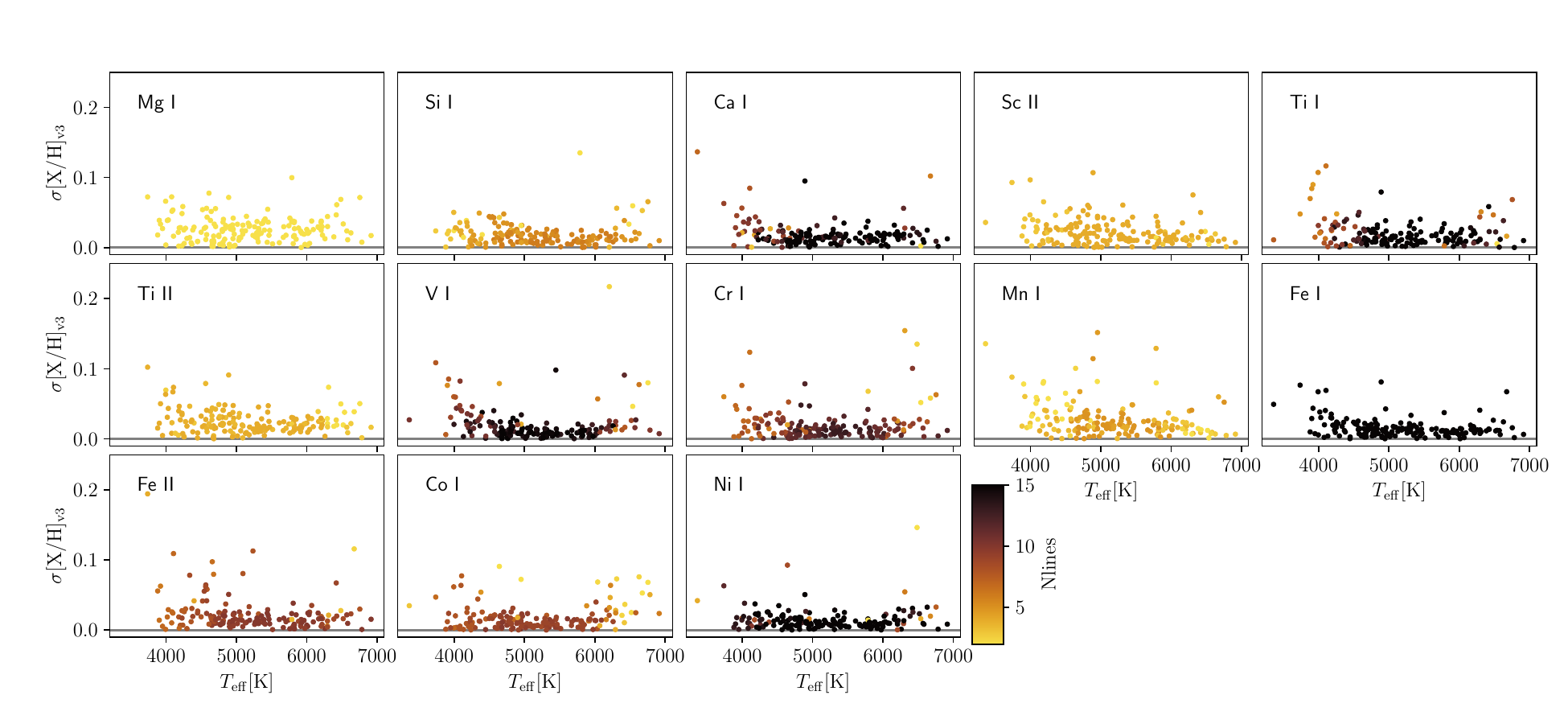}}
\caption{Scatter in abundance obtained among the different spectra of the same star as a function of Teff coloured by the number of lines detected for each star.}
\label{fig:scatter_Nlines}
\end{figure*}

\begin{figure*}[htp]
\centerline{\includegraphics[width=0.7\textwidth]{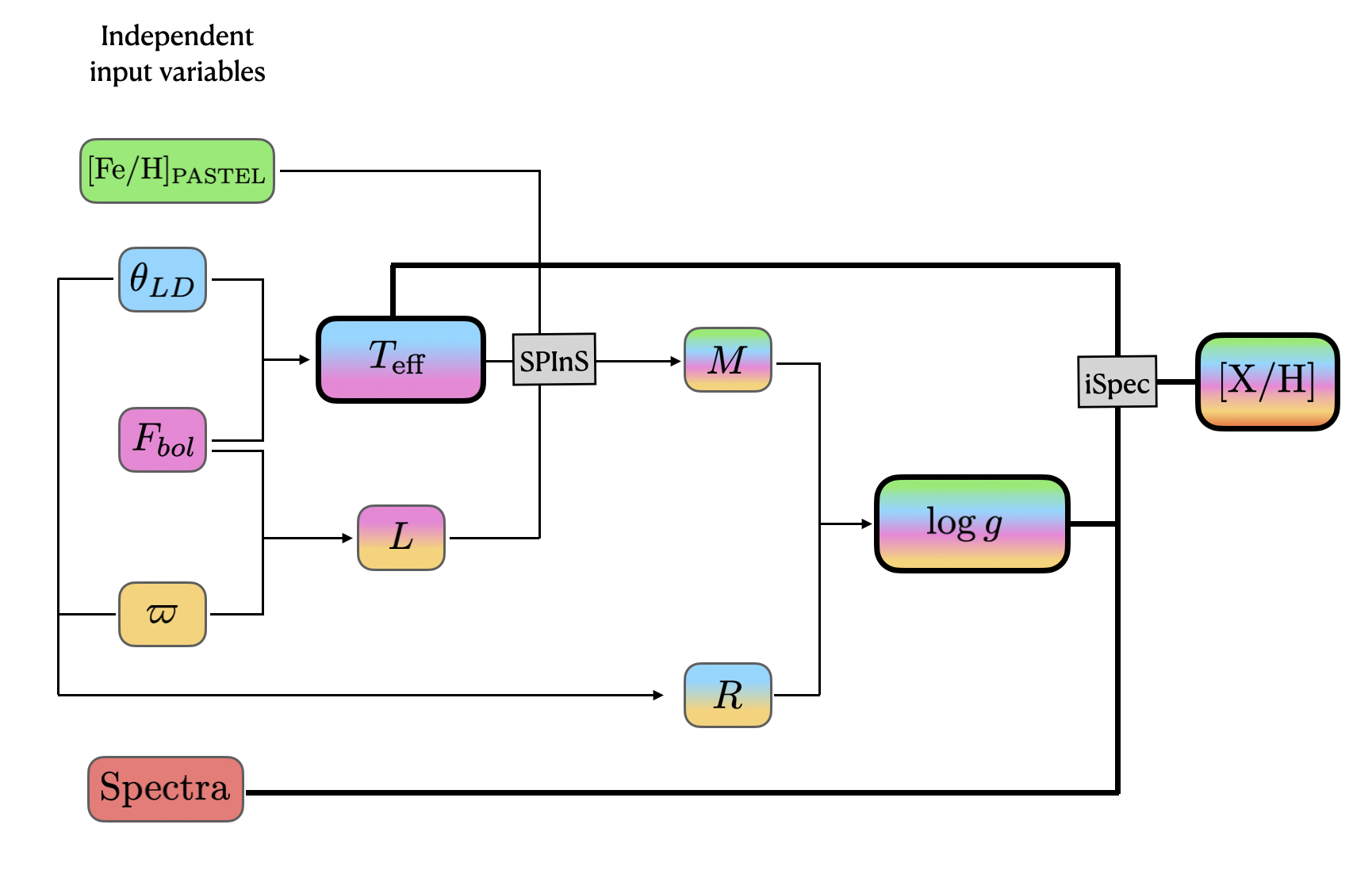}}
\caption{Schematic diagram representing the dependencies of the [X/H] on the various input quantities. Each colour corresponds to one independent quantity.}
\label{fig:diagram}
\end{figure*}

\end{document}